\documentclass{ws-rv9x6}
\usepackage{subfigure}   
\usepackage{ws-rv-thm}   
\usepackage[square]{ws-rv-van}   
\makeindex

\usepackage{lipsum}

\usepackage{xspace}

\begin{document}


\chapter*{End-to-end analysis using image classification}\label{ra_ch1}

\author{Adam Aurisano}
\address{Department of Physics, University of Cincinnati, Cincinnati, Ohio 45221, USA}
\author[A. Aurisano and L. H. Whitehead]{Leigh H. Whitehead}
\address{Cavendish Laboratory, University of Cambridge, Cambridge, CB3 0HE, United Kingdom}

\begin{abstract}
End-to-end analyses of data from high-energy physics experiments using machine and deep learning techniques have emerged in recent years. These analyses use deep learning algorithms to go directly from low-level detector information directly to high-level quantities that classify the interactions. The most popular class of algorithms for these analyses are convolutional neural networks that operate on experimental data formatted as images. End-to-end analyses skip stages of the traditional workflow that includes the reconstruction of particles produced in the interactions, and as such are not limited by efficiency losses and sources of inaccuracy throughout the event reconstruction process. In many cases, deep learning end-to-end analyses have been shown to have significantly increased performance compared to previous state-of-the-art methods.
  \vfill
  \textit{To appear in Artificial Intelligence for High Energy Physics, P.~Calafiura, D.~Rousseau and K.~Terao, eds. (World Scientific Publishing, 2022)}
\end{abstract}
\body

\tableofcontents

\section{Introduction}
End-to-end analyses take their name from the fact that they use a single algorithm that takes raw or low-level detector data as input and outputs high-level physics information for each event. By definition, these analyses skip most, or all, of the traditional workflow for particle physics analyses. Deep learning approaches are a natural choice for these algorithms that can extract features from the input data to perform powerful classifications. Inspired by developments in computer vision and image recognition, a popular choice of algorithm is the 2D convolutional neural network (CNN)~\cite{firstCNN}. CNNs operate on image-like, lattice-structured inputs produced from raw or low-level detector data to generate predictions of physics-level outputs to classify and describe interactions, such as identified particles and overall event-type classifications. 

Section~\ref{sec:traditional} outlines a typical example of the traditional reconstruction and analysis workflow, and Section~\ref{sec:deeplearning} discusses the two primary end-to-end analysis algorithms, CNNs and Graph Neural Networks (GNNs). Uses cases for CNNs are presented for lattice-structured experiments in Section~\ref{sec:latticeCases}, for non-lattice-structured experiments in Section~\ref{sec:heterogeneousCases}, and for time-series data in Section~\ref{sec:1dcnns}.  Section~\ref{sec:gnnCases} describes a use case for end-to-end analysis using GNNs. Section~\ref{sec:networkbehaviour} details methods for probing the behavior of deep neural networks, and Section~\ref{sec:conclusion} provides some concluding remarks.

\section{Traditional Workflow}\label{sec:traditional}
The specifics of each analysis workflow can vary widely between different experiments and sub-disciplines within high-energy physics, but they can typically be broken down into four main steps: low-level reconstruction, particle clustering, particle identification, and event classification. It should be noted that some, or all, of these stages could include machine (and deep) learning aspects, such as boosted decision trees (BDT), neural networks and CNNs in order to make important decisions at key points in the workflow. These stages are discussed briefly below.

\paragraph{Low-level reconstruction}
In this context, low-level reconstruction refers to the finding of signals from the active detector elements, for example the electronic readout channels from a silicon vertex locator in a collider detector, or the readout wires in a liquid argon time projection chamber (LArTPC). These raw signals are processed and converted in some way to produce \emph{hit} objects that form the basis of further event reconstruction. Each hit represents an energy deposit at a given location in space at a specific time.

\paragraph{Particle clustering}
The reconstructed hit objects form the basis of the main particle reconstruction. A set of clustering algorithms are applied to group hits together based on spatial and temporal distance. These clusters are then associated together to build up objects representing each of the individual particles that interacted inside the detector. These objects generally fall into two categories with track- or shower-like topologies. Particles such as muons that lose energy primarily by ionization leave track-like energy deposits in the detectors, whereas particles such as electrons and photons tend to initiate electromagnetic (EM) cascades of particles forming shower-like topologies. The aim is to have a list of fully reconstructed particles at the end of this reconstruction step.

\paragraph{Particle identification}
Once the individual particles have been reconstructed, they can be classified as a specific type of particle. The identification of track-like particles typically includes the use of the energy loss per unit length, $\frac{dE}{dx}$, and the track curvature in a magnetic field for determination of its momentum and the sign of the electromagnetic charge. Topological information in particle cascades, often referred to as \emph{jets} or \emph{showers}, is used to identify particles that produce non-track-like energy deposits.

\paragraph{Event classification}
At this stage the reconstructed interaction contains all of the reconstructed particles with an attached measured particle type. Full events are built from the individual particles and associations are made between the different particles to give the flow of the interaction. Finally, an overall classification of the full physics interaction is given along with important variables that describe the interaction as a whole, for example the energies of the colliding particles.

\section{Deep Learning Approaches}\label{sec:deeplearning}

Image-like data has been common in particle physics since its earliest days.  In particular, the bubble chamber, invented around 1952 by Donald Glaser, was a key detector technology for decades~\cite{Giacomelli:2003fq}.  These detectors dominated the field because they were reliable, fully active, and had high spatial resolution.  All of these properties were key in developing our understanding of hadron properties and electroweak unification. 

\begin{figure}
    \centering
    \includegraphics[width=0.9\textwidth]{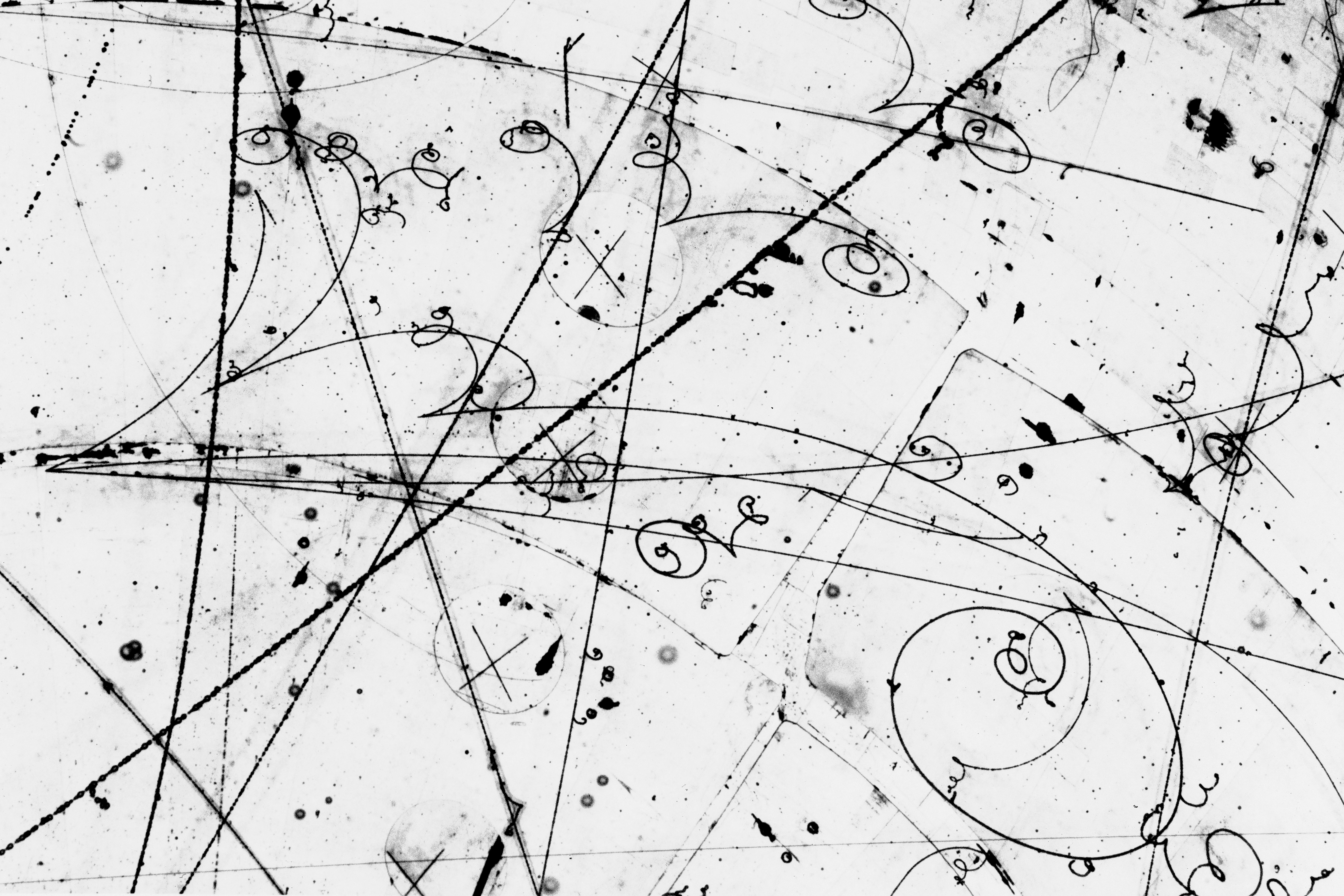}
    \caption{An example of tracks captured by a bubble chamber.  These images were typically manually reconstructed by trained human scanners. Illustration courtesy of Fermilab.}
    \label{fig:bubble_chamber}
\end{figure}

Figure~\ref{fig:bubble_chamber} shows a typical image of tracks recorded by a bubble chamber.  These chambers would be exposed to charged particle beams, and the resulting tracks would be recorded on photographic film.  Trained human scanners reconstructed the decays of particles in the beam manually by visually recognizing and isolating meaningful features like vertices or tracks which the scanners would then physically measure.

\begin{figure}
    \centering
    \begin{tabular}{ccc}
    \includegraphics[width=0.3\textwidth]{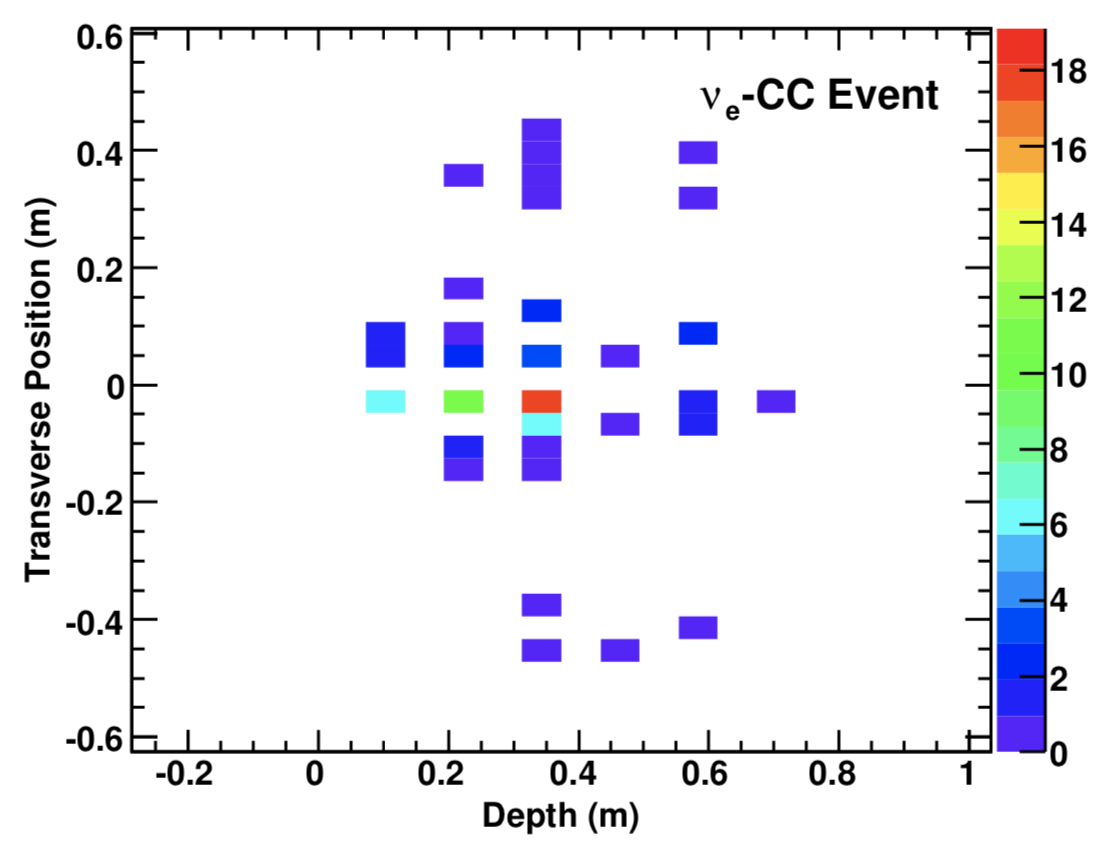} &
    \includegraphics[width=0.3\textwidth]{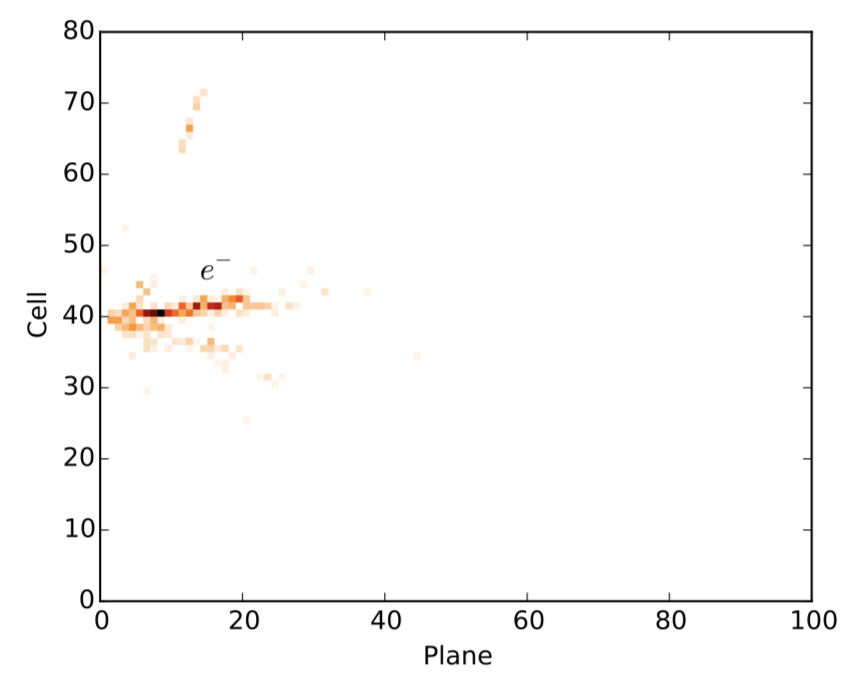} &
    \includegraphics[width=0.3\textwidth]{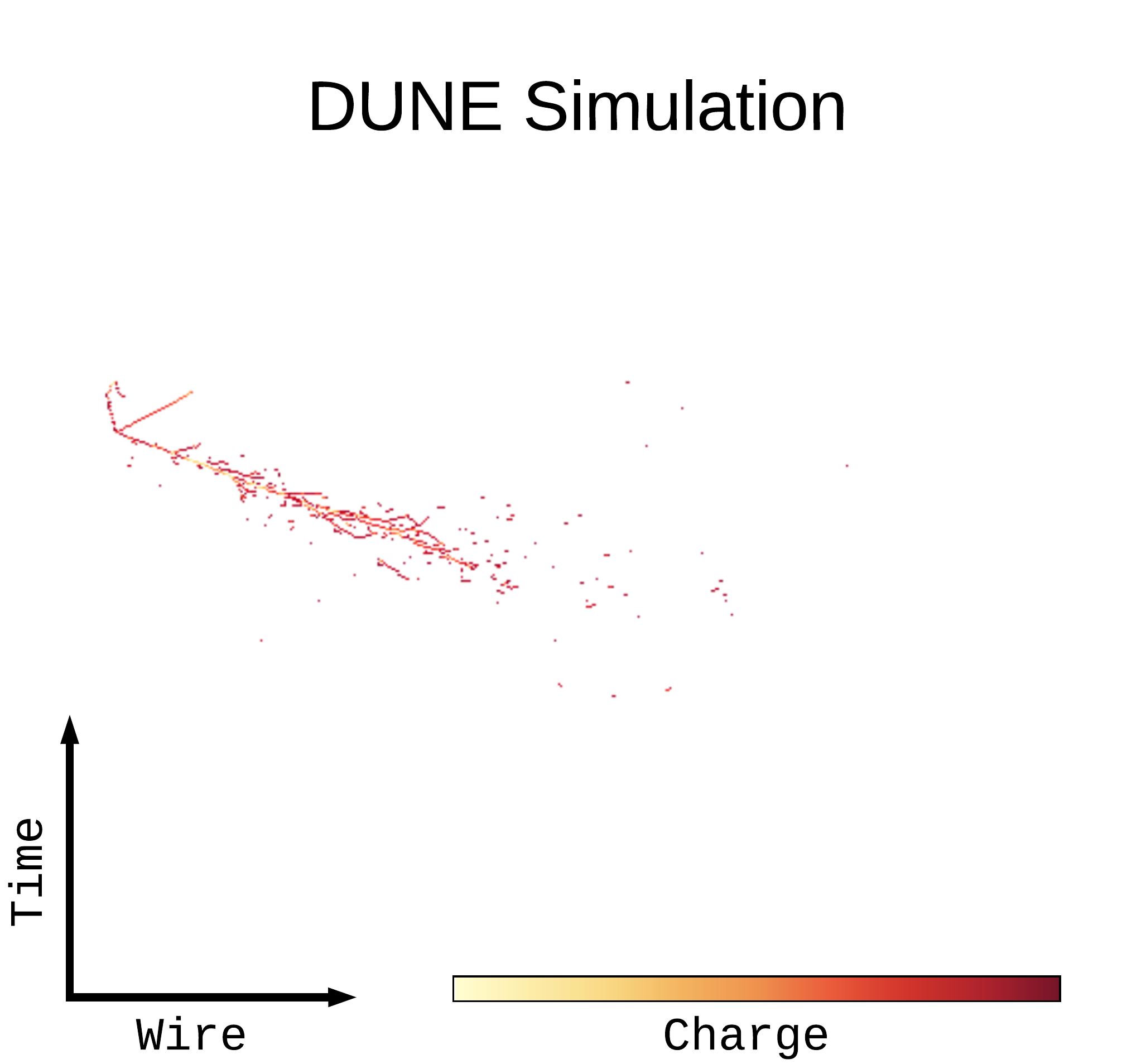}
    \end{tabular}
    \caption{Example CC $\nu_e$ interactions in progressively higher resolution detectors. Left: event display from MINOS from Ref.~\cite{holinThesis}. Center: event display from NOvA from Ref.~\cite{novacvn}. Right: event display from DUNE, adapted from Ref.~\cite{duneCVN}.}
    \label{fig:numu_displays}
\end{figure}

The focus on studying rare interactions required larger detectors with a higher data rate making hand-scanning increasingly impractical. This eventually led to the development of technologies that relied on electronic readout which was automatically reconstructed using techniques described in Section~\ref{sec:traditional}.  While the traditional workflow has proven very successful, current and next-generation high energy physics experiments can provide very fine details of interactions in comparison to previous experiments. To give an example from neutrino physics, charged-current (CC) $\nu_\mu$ interactions consist of a muon with accompanying hadronic activity (nucleons and any number of charged and neutral pions). In the MINOS detectors~\cite{Michael:2008bc}, which are relatively coarse-grained due to the use of thick steel plates between scintillator planes, CC $\nu_{\mu}$ events looked like a long muon track along with a collection of energy deposits from the hadronic activity at the interaction vertex. NOvA~\cite{Ayres:2007tu}, which contains little dead material and is in many ways the successor to the MINOS experiment, provides more detail of the hadronic system and can resolve some of the particles. A next-generation experiment such as DUNE~\cite{duneTDR}, which uses liquid argon time projection technology (LArTPC)~\cite{Rubbia:1977zz}, begins to approach resolutions similar to bubble chambers.  Thus the DUNE detectors will be able to image all of the particles in the hadronic system in fine detail. Figure~\ref{fig:numu_displays} shows example CC $\nu_e$ interactions in each detector and demonstrates graphically the ever-increasing requirements of event reconstruction algorithms to accurately reconstruct the interactions with improving experimental detector resolution. All of the stages of event reconstruction in the traditional workflow are imperfect in terms of both reconstruction efficiency and accuracy. Any mistakes made by the reconstruction algorithms tend to compound through the reconstruction chain and will result in inefficiencies and backgrounds in physics analyses. In addition, each stage of the traditional reconstruction workflow is designed to summarize information about reconstructed features, leading to information loss as summary information from low-level reconstructed objects is combined to form high-level reconstruction objects. Therefore, it is natural to turn to the field of computer vision to find automated approaches to approximate the tasks human scanners performed in previous decades to efficiently use all information collected by the detector.

The inputs to end-to-end deep learning analyses are typically only dependent on the low-level reconstruction and can hence provide powerful analysis-level information without potential errors from the full event reconstruction algorithm chain. However, careful consideration is required for the selection of the training samples for these algorithms to ensure that are not biased due to overfitting or fine-tuning on the training sample. This is particularly important when training on Monte Carlo (MC) simulations based on physics models with associated uncertainties. Two broad categories of deep learning techniques are discussed and demonstrated with examples from various high-energy physics experiments: CNNs and GNNs in Sections~\ref{sec:CNNs}~and~\ref{sec:GNNs}, respectively.

\subsection{Convolutional Neural Networks}\label{sec:CNNs}

The artificial neural network~\cite{rosenblatt1962principles}, also known as a multilayer perceptron (MLP), is a machine learning algorithm characterized by layers of nodes with defined connections between them.  Each node represents a non-linear function of the sum of all input connections, and each connection is associated with a weight parameter which scales the output of one node to become the input of another.  With a suitable selection of weights, usually chosen through a training procedure, an MLP can approximate a wide variety of functions.  

Traditional, or fully connected, MLPs consist of nodes arranged in layers where each node in layer $n-1$ is connected to each node in layer $n$. They have been widely used in high energy physics as selection functions to learn whether or not a given event is signal or background based on a series of input reconstructed quantities.  However, this technique, which sits at the end of the traditional reconstruction workflow, is subject to all potential compounding errors in the event reconstruction, described in Section~\ref{sec:traditional}.

The CNN~\cite{Goodfellow-et-al-2016} is similar to a fully connected MLP except the pattern of connections between nodes on different layers is tightly constrained.  This structure was inspired by studies of the visual cortex of cats and monkeys which determined that cells within the visual cortex were activated by specific illumination patterns on regions of the retina known as receptive fields~\cite{Hubel1959,Hubel1962,Hubel1968}.  The characteristics of the receptive field were observed to vary for different cells within the cortex, leading to the cells being classified as simple, complex, or hypercomplex.  Simple cells are those which are most strongly activated by a static illumination pattern of a specific location, orientation, and shape either on one eye or on corresponding locations on both eyes. Complex cells respond to orientation and shape, but instead of responding to a specific location, they respond to movement of the illumination pattern through a receptive field.  Hypercomplex cells additionally respond to the length of an illumination pattern.  Hubel and Wiesel hypothesized that complex cells received signals from simple cells, and hypercomplex cells received signals from complex cells.  This implies that the mammalian visual cortex analyzes images by using a hierarchical network of cells which have local connections from one layer to the next.  In this way, the brain extracts edge features at the lowest layers, it determines directionality in the middle layers, and it finds extents in the highest layers. 

CNNs were first used in the 1980s to identify handwritten digits~\cite{firstCNN}, but they did not become widespread until 2012 with the success of AlexNet~\cite{alexnet} in identifying images in the ImageNet challenge~\cite{Russakovsky2015}. The dramatic success of AlexNet has since produced a proliferation of CNN architectures that have improved image classification to super-human levels; however, despite the diversity in architectures, all CNNs share some common structures, which we will detail below.  

The structure of CNNs begins from the insight that images can be interpreted as an array of numbers of size $h\times w \times c$, where $h$ and $w$ are the height and width of the image, respectively, in pixels, and $c$ is the number of channels (also known as the depth).  For greyscale images, $c=1$, while for color images, $c=3$.  The value of each array element represents the intensity of the light at the corresponding location and channel.

This array is then passed through a series of layers which perform local operations under the assumption that pixels close to each other are likely to be semantically related.  Each layer produces a series of $h \times w \times c$ outputs known as feature maps, with potentially changed sizes of $h$, $w$, and $c$.  As feature maps are passed through layers, features represent increasingly more global information as local information from larger regions is combined.  The full column of layers learns to automatically extract high-level features which replace the use of hand-crafted features that are typically used in traditional MLPs. The features produced by the final layer are fed into a single layer of a traditional MLP to produce the outputs which approximate the function the network was trained to learn.

Many types of layers have been developed for use in CNNs, but the most important types are convolutional layers and pooling layers.  All CNNs contain convolutional layers which directly mimic the concept behind the simple cell in the visual cortex.  A simple cell receives inputs from a local region of the retina and weights each input according to a particular excitatory or inhibitory pattern.  Convolutional operators mimic this by taking the dot product of the values in a local region of a feature map with a matrix of learnable weights known as a convolutional kernel.  This operation is repeated across the entire feature map to produce an output feature map.  This operation is closely related to the discrete convolution.

The choice to use the same convolutional kernel to extract features across the full input produces two properties that are responsible for much of the power of CNNs.  First, learning a single kernel per output feature map dramatically reduces the number of free parameters learned by the network compared to fully connected layers in a traditional MLP.  This makes CNNs easier to train and less susceptible to overtraining.  Second, applying the same operation at every local region makes the layer equivariant to translation.  That is, if an object is translated in an image, the same features will be produced, just translated within the output feature map.  If this property is maintained throughout the network, the efficiency for detecting if a particular object is present in an image will not depend on the exact location of the object in the image.  However, scale, rotation, and the relative positions of objects within the image can still be important.

Pooling layers are optional in CNNs, but they extremely common. They apply a pooling operator which combines features from a local patch in an irreversible way.  The most commonly used pooling operator is max pooling, which outputs the maximum value of the features in a local patch. For example, a $2\times2$ max pooling layer downsamples $2\times2$ input pixels to a single pixel with the value of the highest-valued input pixel. This has the effect of removing low significance information and imposing an invariance to small translations. The invariance property reduces the sensitivity of the network to exact positions, including relative positions.
 
\subsection{Graph Neural Networks}\label{sec:GNNs}
There are many situations where representing experimental data as an image is not a natural or convenient method. For example, experiments with complex geometries may require numerous projections of the data to produce 2D images. Even in experiments where images provide a good data representation, there are cases where only a small fraction of the detector elements are activated for a given event, resulting in images with many empty pixels (see many of the images discussed in Sec.~\ref{sec:CNNs}). This does not necessarily present a problem, but a lot of time can be spent performing convolutions on pixels with zero values, which always results in a zero result regardless of the filter applied. 

In the case of complex geometries, a more natural representation may involve considering each detector element as a point in 3D space. Furthermore, considering only those detector elements with a measured signal on an event-by-event basis will provide a more efficient approach for sparse data. A data structure that copes well with these requirements is a graph
\begin{equation}
    G = (V,E),
\end{equation}
where: $V$ are a set of vertices, also known, and henceforth referred to, as \emph{nodes}; and $E$ are a set of \emph{edges}. Edges are defined as connections between nodes such that edge $e_{ji}$ links node $v_j$ to node $v_i$. This is an example of a \emph{directional} edge since it points from node $v_j$ to node $v_i$. Edges can also be \emph{undirected}, in which case $e_{ji} = e_{ij}$ and the link is reciprocated. Each node has a number of \emph{features} associated to it that describe its properties.

To form graphs from high energy physics data, each detector element with a measured energy deposit is added as a graph node. Each node has a number of features, which could include information such as the position of the detector element and the amount of deposited energy. The ability to associate multiple features with each node provides an easy way to incorporate more information into a GNN beyond just position and charge. The edges that link the nodes can be defined in a number of ways, for example using the adjacency of nodes to their neighbors. Each interaction is therefore represented as a connected graph containing all of the recorded energy deposits in the detector.

Graph Neural Networks~\cite{SperdutiFirstGNN,Gori2005ANM,ScarselliGNN,bruna2013spectral,henaff2015deep} (GNNs) are neural networks that operate on graphs. Depending on the specific classification task, the GNN can classify nodes, edges or the entire graph. There is a large variety of GNN architectures~\cite{Wu_2020}, but they typically use graph-based convolutions to aggregate the features of a node and its neighbors. 
 
\subsection{Network Optimization}\label{sec:opt} 

Convolutional and graph neural networks, like all machine learning algorithms, need to undergo a training procedure, and the performance of the final algorithm is highly dependent on the quality of the training. The choice of training samples, typically from simulation, is a key consideration to ensure that the algorithm generalizes well, meaning that it performs similarly on data not included in the training sample. There are many other important factors, including the choice of optimizer used to find the minimum of the loss function, and the network \emph{hyperparameters}.

\paragraph{Training samples}The choice of training sample is very important for CNNs and GNNs. These networks typically have of the order of millions of parameters and therefore need large training samples to be successfully trained and optimized. The vast majority of networks used in end-to-end analysis are trained using simulated data events and the associated truth information is used to provide the target labels. It is very important to ensure that the training sample covers the entire range of possible interactions that could be seen in the samples that the network will be used to classify.

 \paragraph{Optimizers}In machine learning, the training process minimizes a loss function that describes how close the prediction is to the true value(s). The loss function exists in a very high dimensional space and varies as a function of the trainable parameters that form the network model. Gradient descent is the general method for finding the (local) minimum of the loss function, whereby gradients are calculated with respect to each parameter and then parameter values are updated using the negative of the gradients multiplied by a factor called the \emph{learning rate}. Many optimizers are variants of Stochastic Gradient Descent~\cite{robbins1951} (SGD), a method where the parameter values are updated after each training example (or more commonly, mini-batch of examples). There are a number of different optimizers that are extensions to the standard SGD algorithm that aim to improve performance and ensure robustness, such as ADADELTA~\cite{zeiler2012adadelta}, RMSProp~\cite{rmsprop}, Adam~\cite{kingma2017adam}, etc. For a more detailed discussion of optimizers, see, for example, Ref.~\cite{optimizersummary}.
 
 \paragraph{Hyperparameters}Parameters that can not be optimized during training are known as hyperparameters. These are usually either parameters controlling the structure of the network itself, like the number and type of layers, or controlling the behavior of the optimizer.  One of the most important hyperparameters of the latter type is the learning rate.  If it is too large then the optimizer can fail to find a minimum in the loss function, but if it is too small then the optimizer can get stuck in a local (and possibly shallow) minimum. More complex approaches involve decaying the learning rate as a function of the training time in order to avoid local minima and fall into the bottom of a deep (or hopefully global) minimum. 
 


\section{Convolutional Neural Networks in Lattice-Structured Experiments}\label{sec:latticeCases}

Due to the low interaction rate of neutrinos, neutrino detectors are typically large, and since neutrinos are equally likely to interact anywhere within the volume of the detector, neutrino detectors are typically homogeneous. Therefore, many neutrino detectors produce data which can be easily reinterpreted as images. As such, the first uses of CNNs to perform end-to-end analyses in particle physics occurred at 
neutrino experiments.  

In addition, similar detector technologies are used in neutrinoless double beta decay and nuclear physics experiments.  The lattice-structured geometries in these experiments lead to commonalities in the approaches used.

\paragraph{Event classification in the NOvA Experiment}

The NOvA experiment~\cite{Ayres:2007tu} is a long-baseline neutrino experiment designed to measure $\nu_{\mu}$ disappearance and $\nu_{e}$ appearance in a beam originally composed of mostly $\nu_{\mu}$. NOvA measures the flavor content and energy spectrum of the neutrino beam at a near and far location using two functionally identical detectors located on the surface and composed of layers of alternating vertical and horizontal liquid-scintillator-filled PVC cells.  The alternating structure provides two orthogonal views of the 3D pattern of energy deposits produced by charged particles traversing the detector projected on the $x-z$ and $y-z$ planes.

\begin{figure}
    \centering
    \begin{tabular}{cc}
    \includegraphics[width=0.45\textwidth]{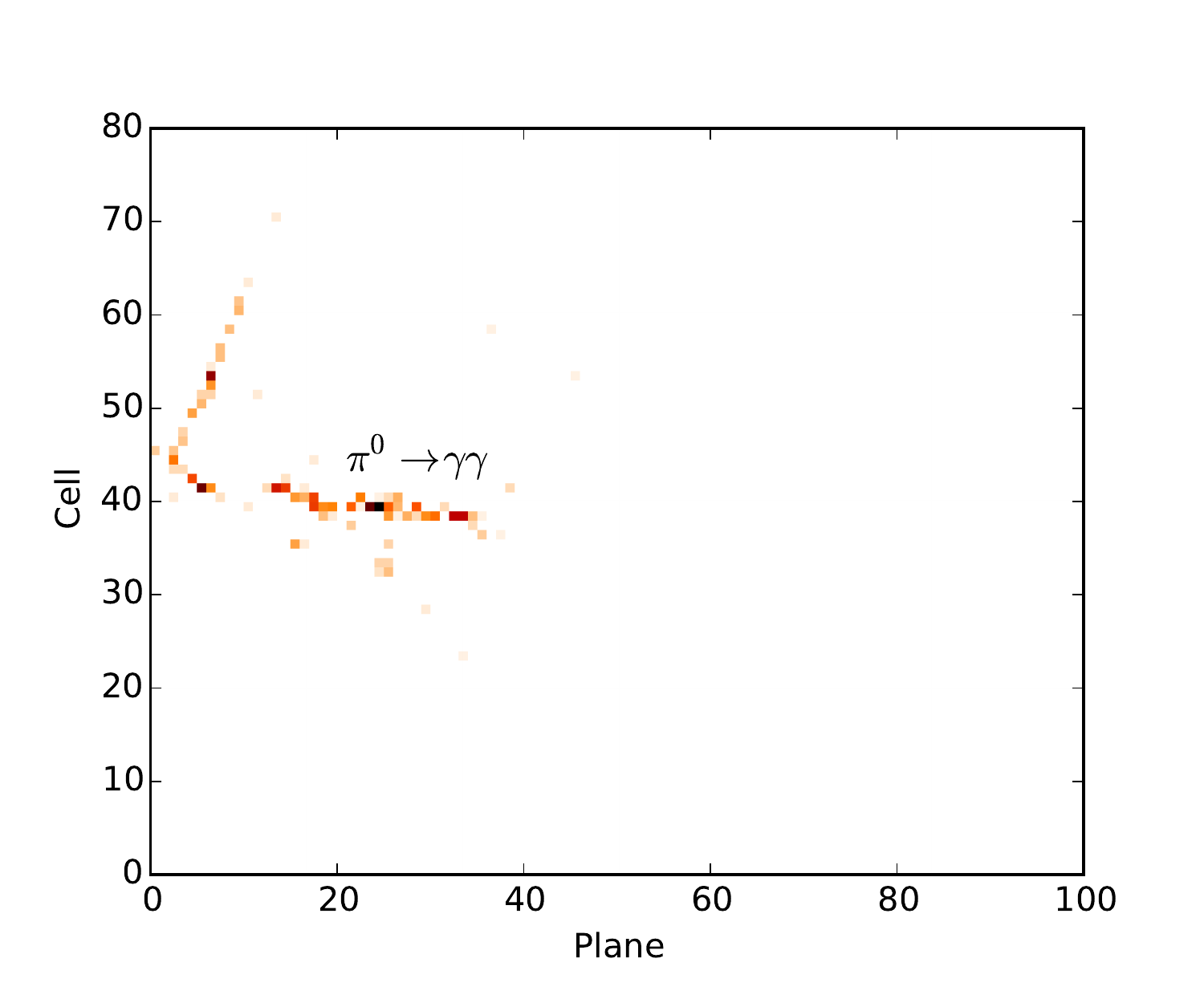} & 
    \includegraphics[width=0.45\textwidth]{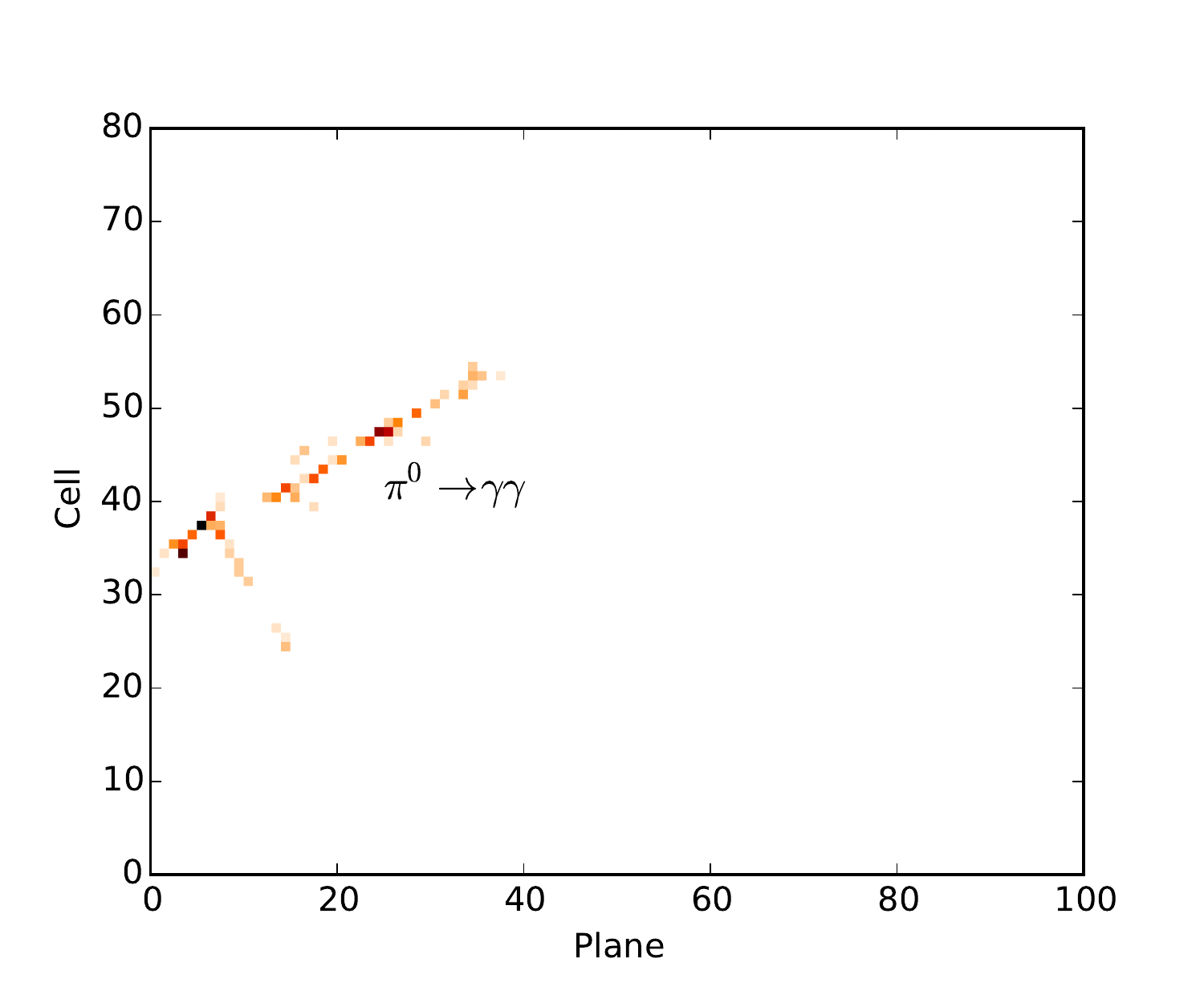}
    \end{tabular}
    \caption{Example of a pair of input pixel maps for a neutral current interaction containing an electromagnetic shower produced by a $\pi^0$. Each view corresponds to either a projection on the $x-z$ plane (left) or the $y-z$ plane (right). Figures reproduced from Ref.~\cite{novacvn}}
    \label{fig:nova_pixel_map}
\end{figure}

To perform oscillation analyses, it is critical to be able to separate events into charged- and neutral-current (NC) interactions, and in the CC case, events must further be separated according to flavor.  Since NOvA is on the surface, the cosmic ray flux is large, so it is also necessary to distinguish between cosmic ray and neutrino events.  A CNN algorithm was developed to achieve this separation~\cite{novacvn}.  The input to the CNN consists one image of 100 planes $\times$ 80 cells for both views.  These smaller images are extracted from the much larger images representing the full detector by performing a clustering of energy deposits in space and time.  This clustering is the only reconstruction performed on NOvA data prior to feeding it into the CNN.  An example of these inputs for an NC event is shown in Figure~\ref{fig:nova_pixel_map}.

The NOvA network is based on a modified GoogLeNet architecture~\cite{Szegedy2015}. The hallmark of this network is a network-in-network design~\cite{Lin2014NetworkIN} where miniature CNNs consisting of several convolutional layers operating in parallel with a variety of kernel sizes form a repeatable ``Inception module''. Feature maps from each parallel branch in the module are merged together and resampled using a $1 \times 1$  convolutional layer.

\begin{figure}
    \centering
    \begin{tabular}{cc}
    \includegraphics[width=0.45\textwidth]{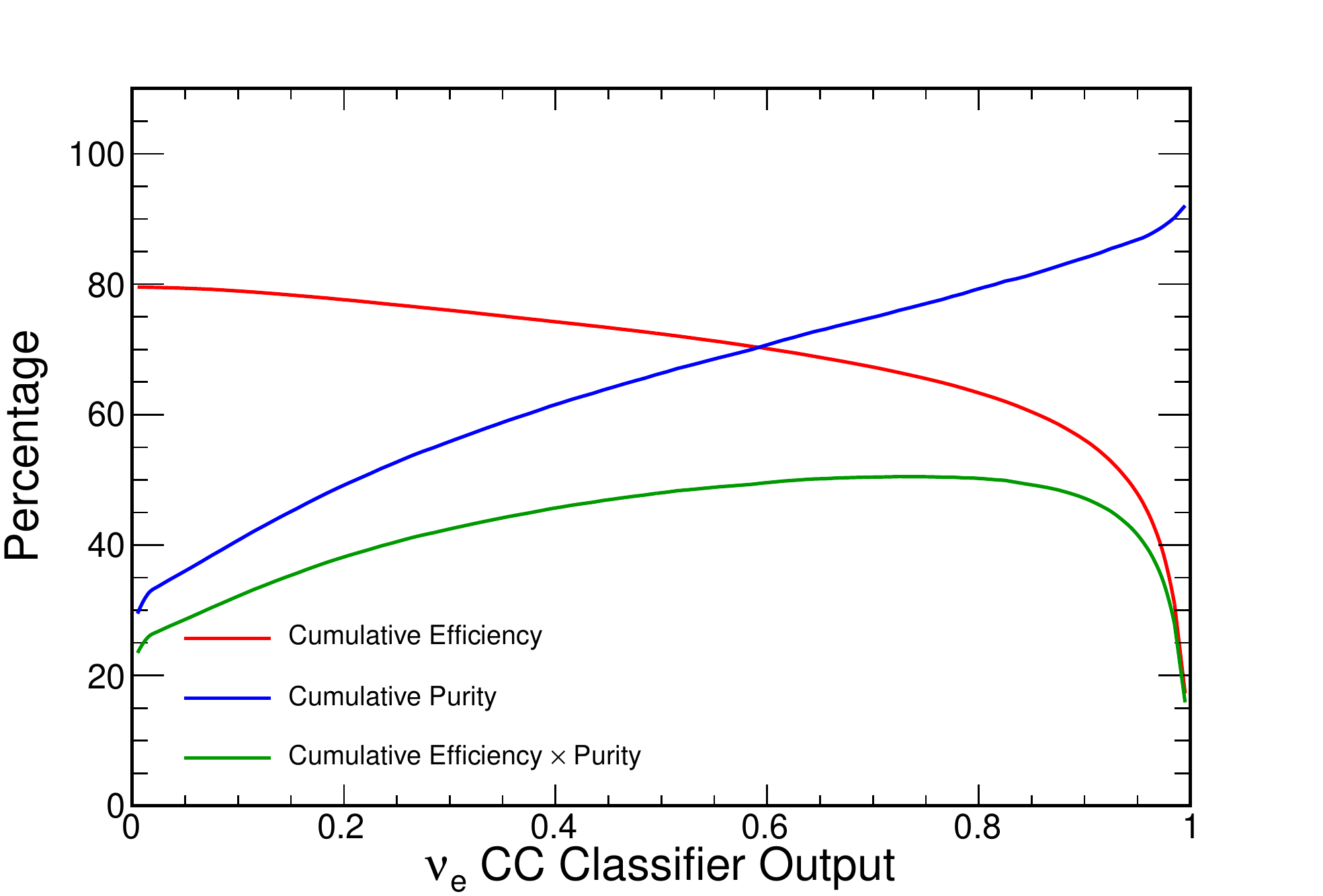} & 
    \includegraphics[width=0.45\textwidth]{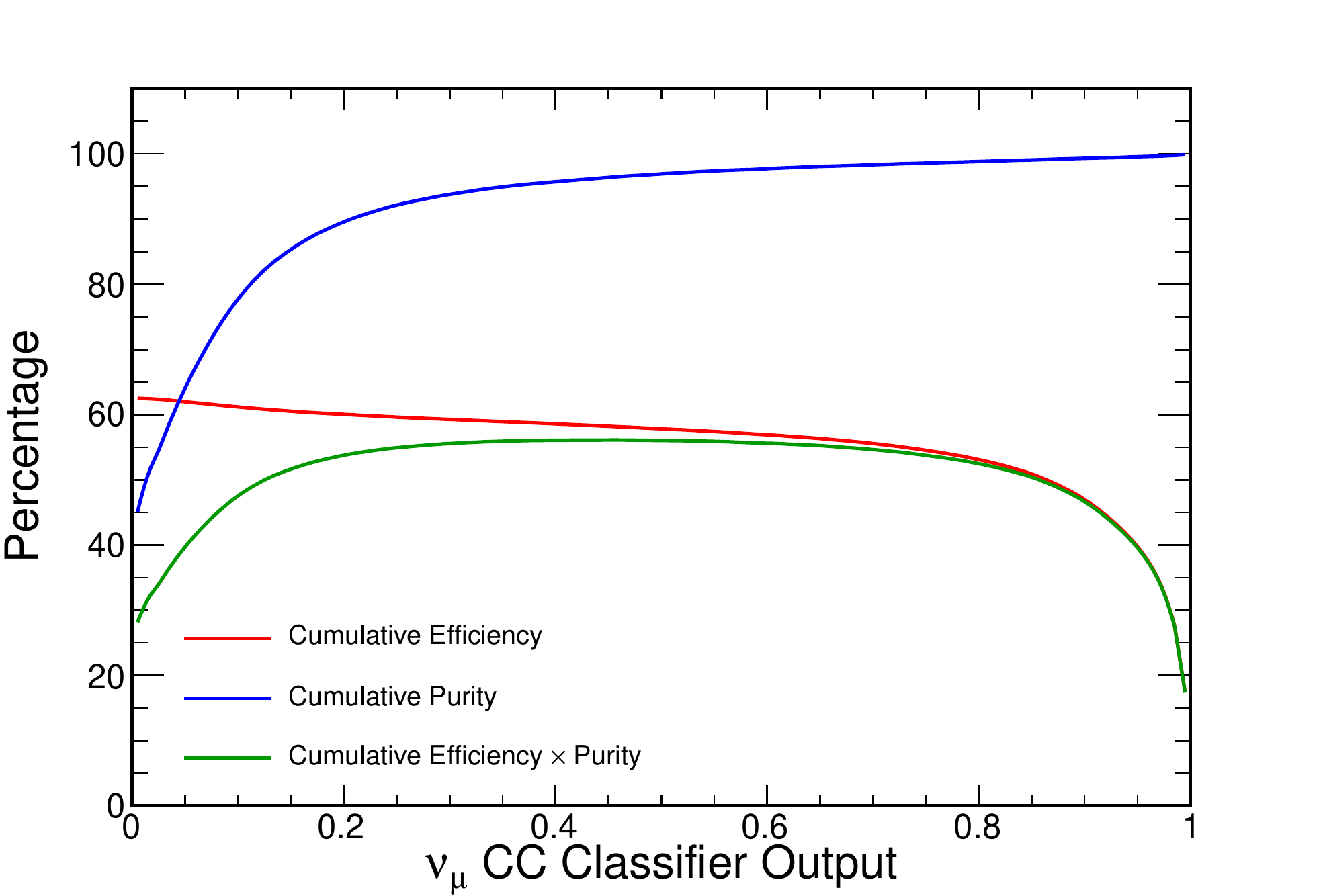}
    \end{tabular}
    \caption{Classifier efficiency (red), purity (blue), and their product (green) for CC $\nu_{e}$ (left) and CC $\nu_{\mu}$ (right) interactions.   Figures reproduced from Ref.~\cite{novacvn}.}
    \label{fig:nova_eff_pur}
\end{figure}

Since the input NOvA pixel maps consist of two views of the same data sharing one common axis and one different axis there is no guarantee that the same pixel location on the two views are physically correlated.  Therefore, each view is processed separately by two branches of the CNN containing three inception modules.  After this stage, the resulting feature maps are sufficiently abstract that they can be concatenated and passed through one final inception module.  Separate outputs of the network predict if an event is a cosmic ray, or if it is a neutrino, its flavor and interaction type. Figure~\ref{fig:nova_eff_pur} shows the performance of the CC $\nu_{e}$ and CC $\nu_{\mu}$ classification outputs.  The network produces a 40\% increase in CC $\nu_{e}$ selection efficiency over previously used traditional selection techniques with no loss in purity.

This network is the first use of a CNN in a published particle physics analysis~\cite{Adamson:2017gxd,Adamson:2017zcg}.  Due to its versatility, the network presented here, as well as subsequent improved networks, has formed the basis of all NOvA oscillation analyses.

\paragraph{Event localization and classification in the MicroBooNE experiment}

The MicroBooNE detector~\cite{Acciarri:2016smi} is a LArTPC located on the surface on the Fermilab campus with three wire readout planes that collect ionization charge liberated by charged particles traversing the detector medium, and a photon detection system to measure scintillation light. CNN-based algorithms were used, for the first time in a LArTPC experiment, to perform classification of cosmic ray and neutrino interactions~\cite{Acciarri:2016ryt}. A number of studies were performed, two of which are discussed below.

\begin{figure}
    \centering
    \begin{tabular}{cc}
    \includegraphics[width=0.45\textwidth]{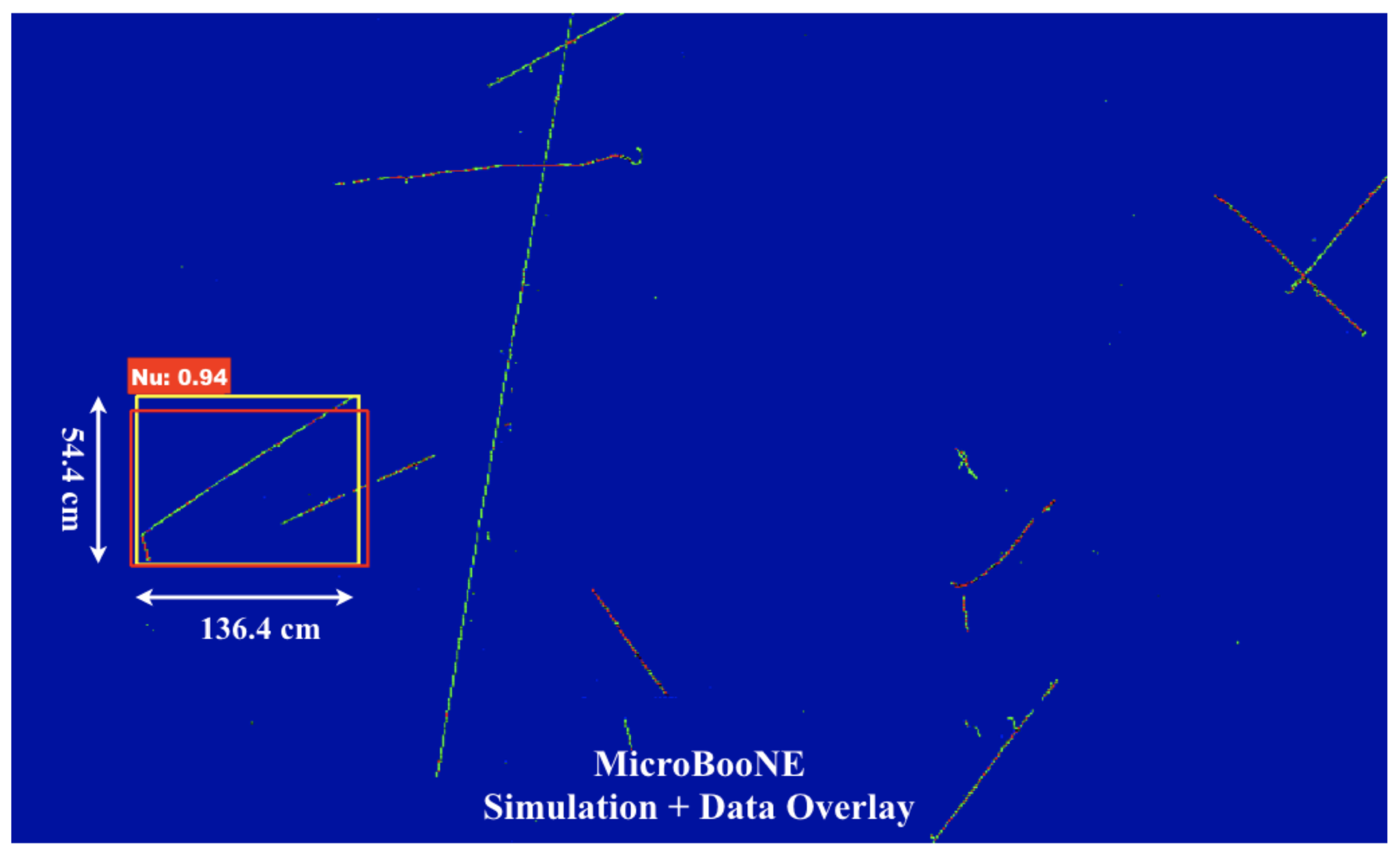} & 
    \includegraphics[width=0.45\textwidth]{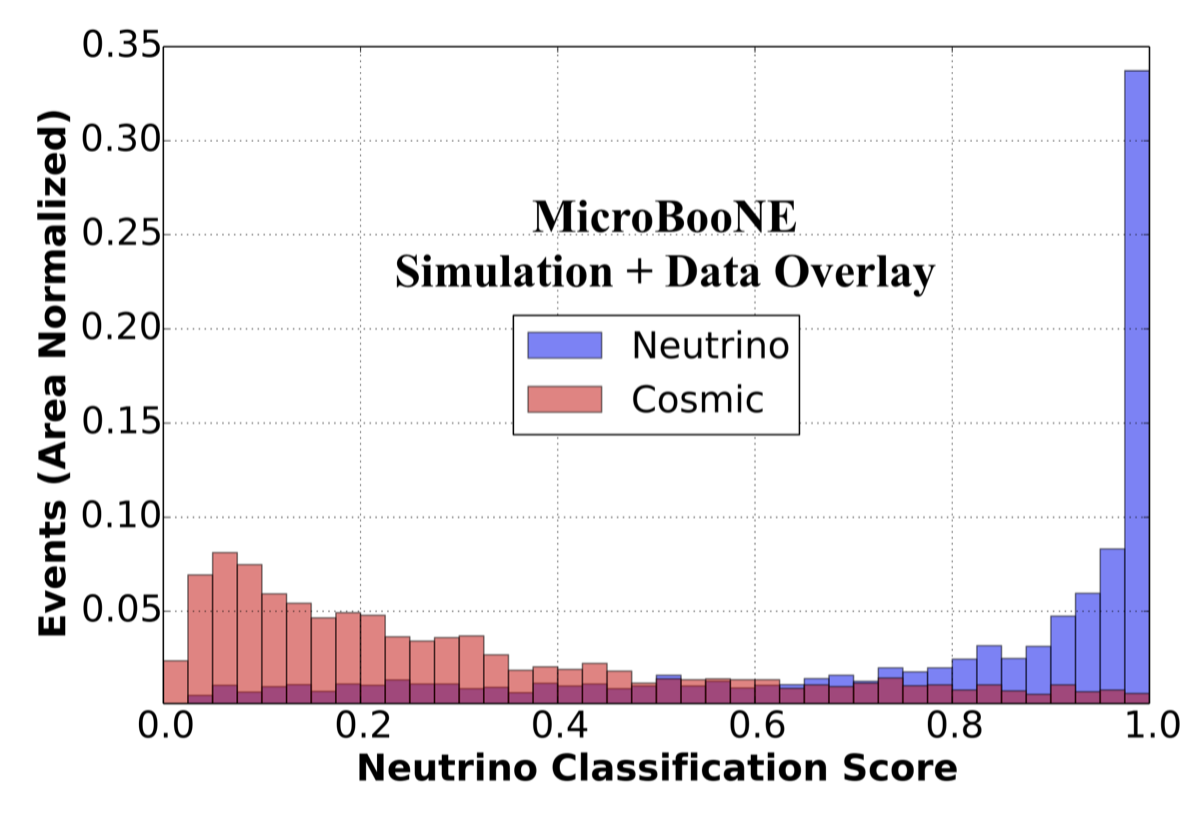}
    \end{tabular}
    \caption{Left: identified neutrino interaction with a bounding box compared between prediction (red) and truth (yellow). The true bounding box is defined as the smallest region containing all true charge depositions from the simulation, and the prediction is the output of the CNN. Right: the neutrino classification score for cosmic and neutrino interactions. Figures reproduced from Ref.~\cite{Acciarri:2016ryt}}
    \label{fig:MicroBooNE_bounding_box}
\end{figure}

The first CNN algorithm using the charge information from a single readout plane was developed to perform the event classification and find the bounding box containing the neutrino interaction. The network uses a hybrid architecture based on AlexNet~\cite{alexnet} and Faster R-CNN~\cite{fasterRCNN}. An example of a correctly classified CC $\nu_\mu$ interaction is shown in Figure~\ref{fig:MicroBooNE_bounding_box} with a large overlap of the true (yellow) and predicted (red) bounding boxes, and the distribution of the neutrino classification score is shown on the right.

\begin{figure}
    \centering
    \includegraphics[width=0.6\textwidth]{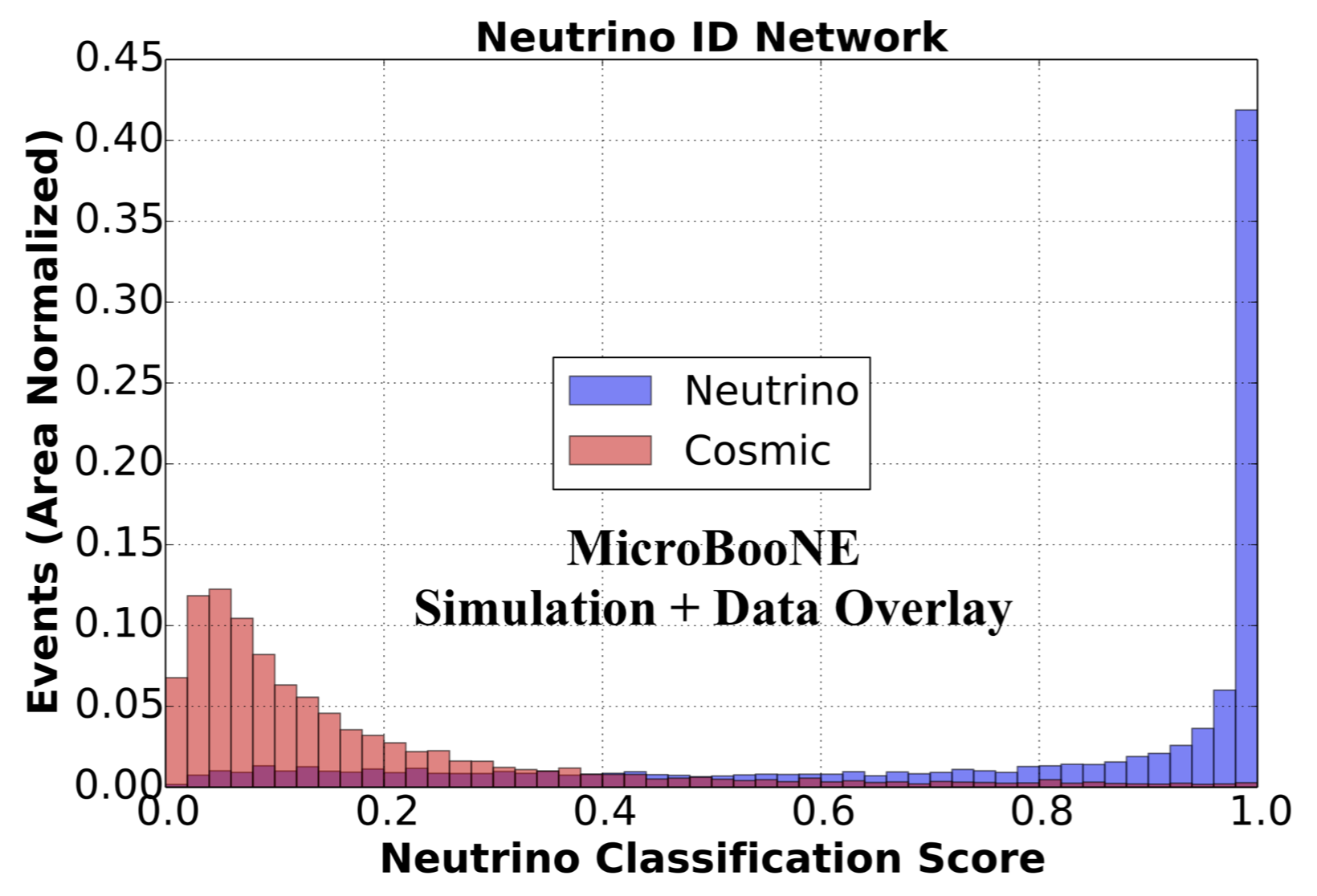}
    \caption{The neutrino classifier distribution for cosmic and neutrino interactions using all detector information. Figure reproduced from Ref.~\cite{Acciarri:2016ryt}}
    \label{fig:MicroBooNE_nu_performance}
\end{figure}

The second CNN algorithm used all three readout views plus the photon detector system. The input for the network consists of a $768\times768$ pixel depth-12 image, formed from three depth four images (one for each of the three readout views). The components of the depth four images are the following features for a given wire and time: the deposited charge; a binary map of charge deposits consistent with minimum ionising particles, such as muons and pions; a binary map of charge deposits consistent with heavily ionising particles, such as protons; and a deposited charge map weighted by the distance of the charge detection point from the averaged light collection point from the photon detectors. The last of these images is used to help the CNN find the most important region of the detector where the neutrino interaction is most likely to have occurred. The network architecture was based on the ResNet~\cite{He-et-al-2015-deep}, using three convolutional layers followed by nine ResNet modules, with two final outputs that gives scores for the interaction to be a cosmic ray or neutrino event. Figure~\ref{fig:MicroBooNE_nu_performance} shows the performance of the classifier for simulated neutrino interactions overlaid with cosmic rays from data. A comparison with the distribution on the right hand side of Figure~\ref{fig:MicroBooNE_bounding_box} shows that including of all the detector information gives a significant improvement in the classification, as expected.

\paragraph{Event and particle content classification in the DUNE experiment}

\begin{figure}
    \centering
    \begin{tabular}{ccc}
    \includegraphics[width=0.32\textwidth]{images/cvn_nue_view0.pdf} &
    \includegraphics[width=0.32\textwidth]{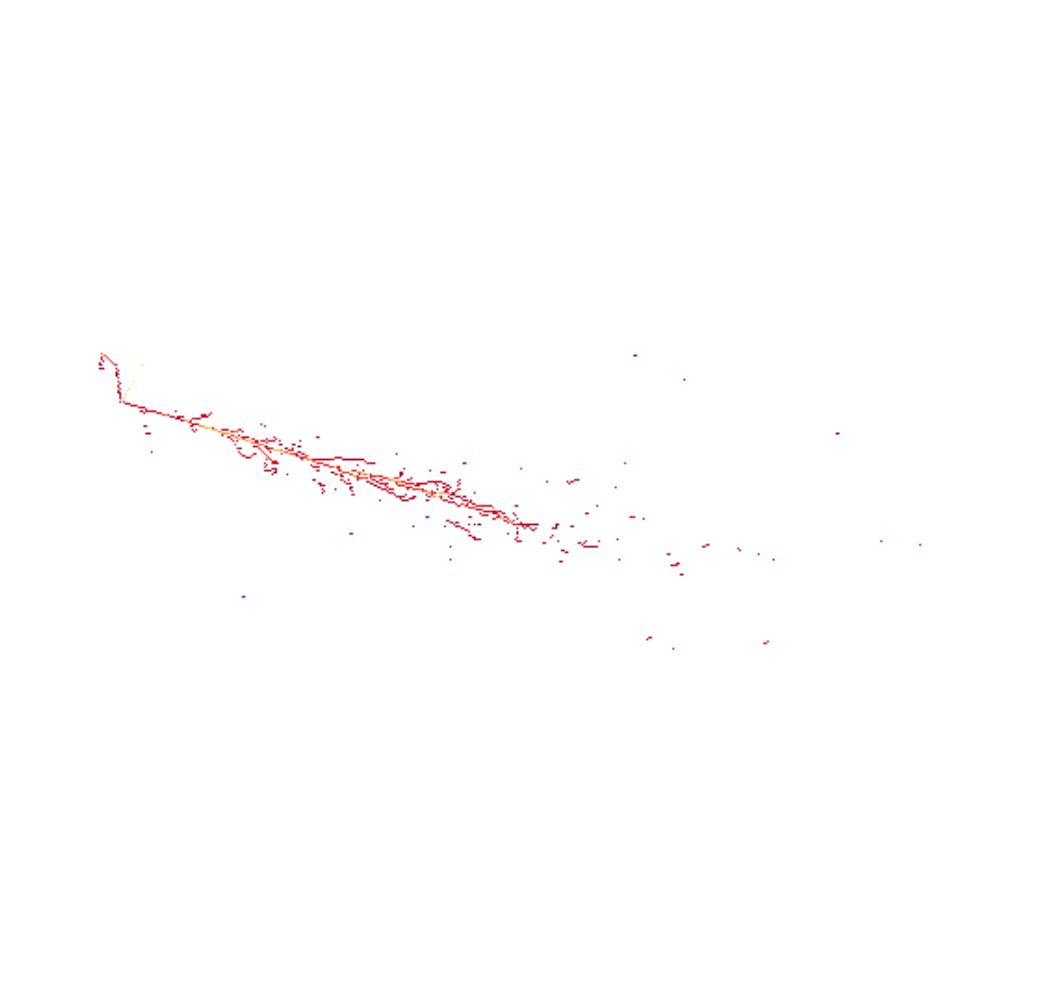} &
    \includegraphics[width=0.32\textwidth]{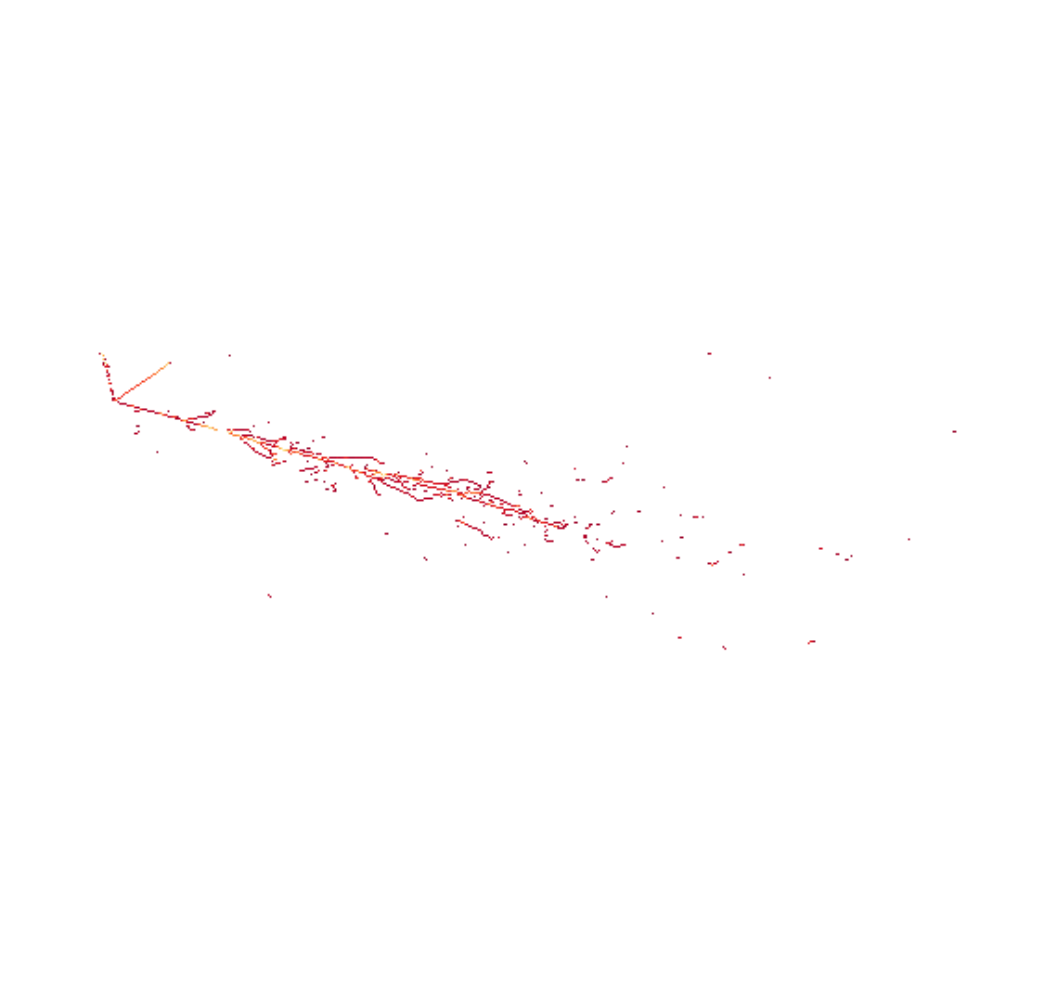}
    \end{tabular}
    \caption{An example of a simulated CC $\nu_e$ interaction in the DUNE far detector, shown in each of the three independent readout views. Figure adapted from Ref.~\cite{duneCVN}.}
    \label{fig:dune_event}
\end{figure}

The Deep Underground Neutrino Experiment (DUNE)~\cite{duneTDR} is a next-generation long-baseline neutrino oscillation experiment. The detectors will use LArTPC technology with three wire readout planes. The data from each of these three readout planes can be visualized as a 2D image with coordinates of wire number and time, as shown for a CC $\nu_e$ interaction in Figure~\ref{fig:dune_event}, where the time coordinate is common between the three images. The 500$\times$500 pixel images are cropped around the neutrino interactions. The pixel values represent the reconstructed charge measured on a given wire at a given time.  The DUNE CNN algorithm~\cite{duneCVN}, also known as the CVN, has an architecture based on the SE-ResNet-34~\cite{He-et-al-2015-deep,He-et-al-2016-identity,Hu-et-al-2017-squeeze}. The initial layers of the network are divided into three branches, one for each of the input images, and seven convolutional layers are applied before the branches are merged together. The final fully-connected layer has a number of different outputs but the primary one is designed to identify the type of neutrino interaction. A number of the other outputs from the CVN provide numbers of different final-state particles visible in the neutrino interactions, including protons, charged pions and neutral pions. 

\begin{figure}
    \centering
    \begin{tabular}{cc}
        \includegraphics[width=0.47\textwidth]{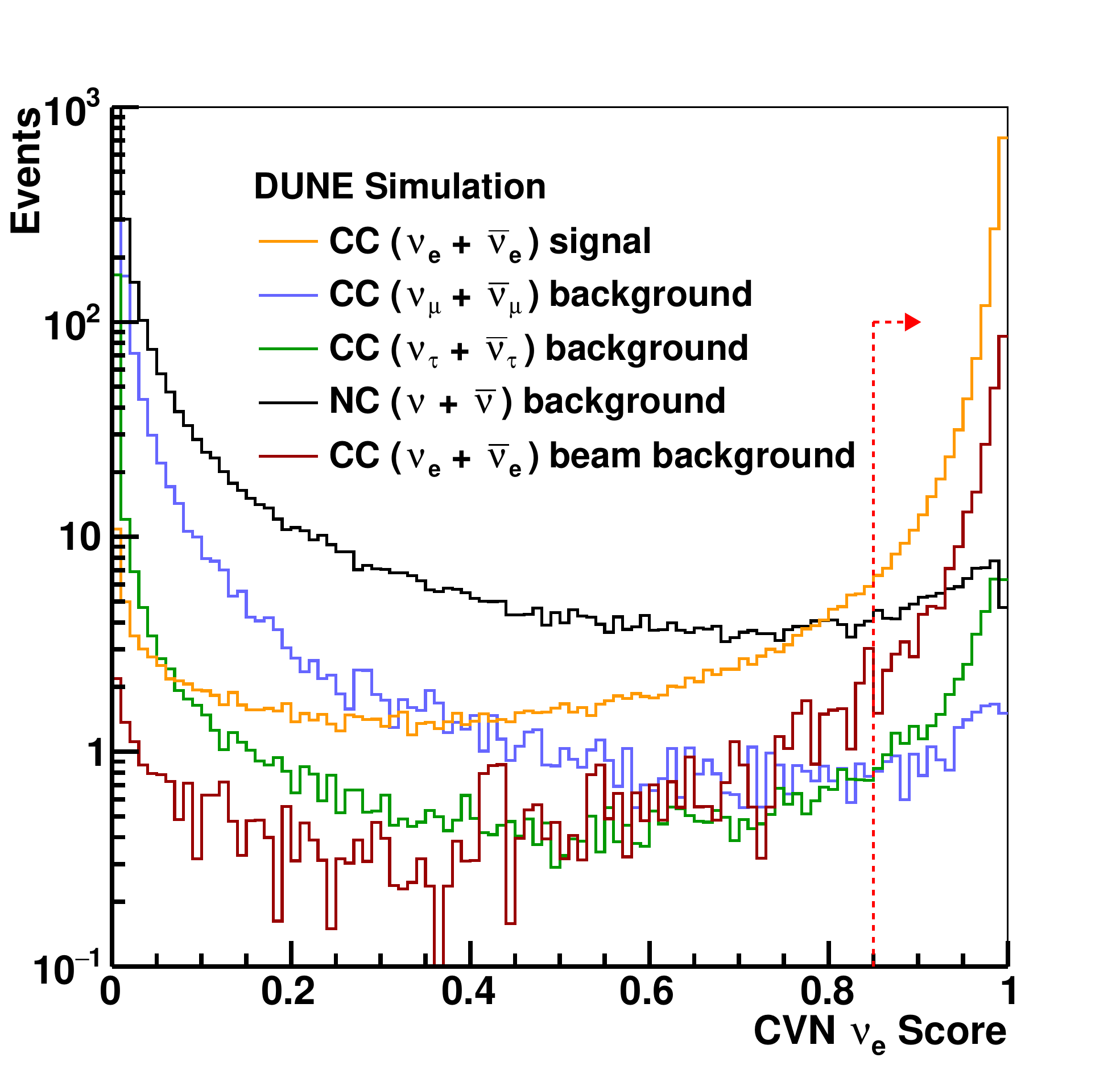} &
        \includegraphics[width=0.47\textwidth]{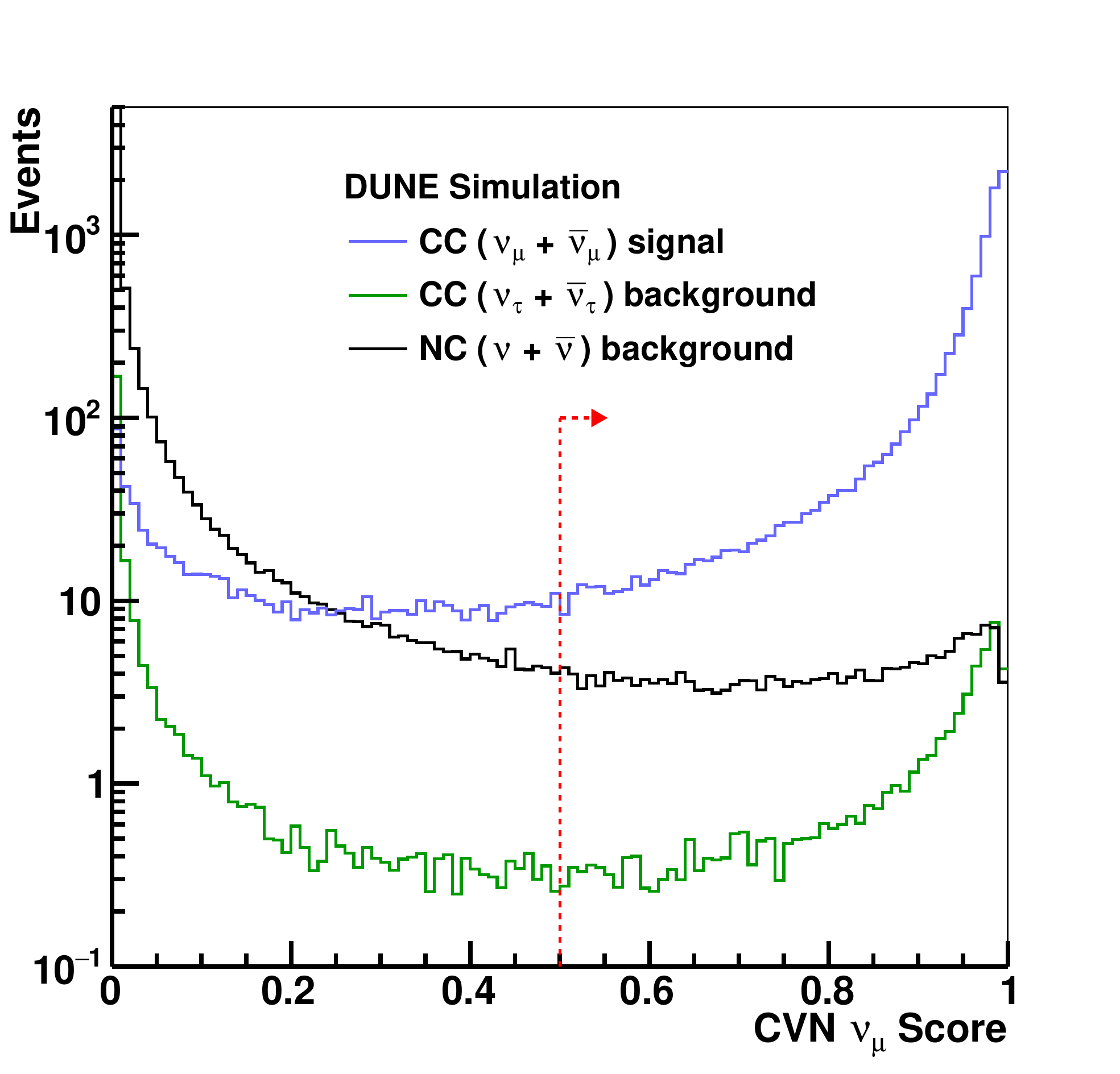}
    \end{tabular}
    \caption{The output score distributions from the DUNE CVN for the CC $\nu_e$ (left) and CC $\nu_\mu$ (right) hypotheses, shown for the various neutrino flux components. The red arrows correspond to the cut values used in the DUNE event selections~\cite{duneLBL}. Figure reproduced from Ref.~\cite{duneCVN}.}
    \label{fig:dune_cvn_scores}
\end{figure}

The neutrino flavor output of the CVN contains four nodes and returns a score for the event to originate from one of four broad categories: CC $\nu_\mu$, CC $\nu_e$, CC $\nu_\tau$ or NC. These output scores provide very powerful neutrino event classification, as shown by the CC $\nu_e$ score and CC $\nu_\mu$ score distributions on the left and right of Figure~\ref{fig:dune_cvn_scores}, respectively. These scores are used to produce event selections for the neutrino oscillation sensitvities described in detail in Ref.~\cite{duneLBL}. CC $\nu_e$ ($\bar{\nu}_e$ interactions are selected with over 90\% (95\%) efficiency.

\begin{figure}
    \centering
    \includegraphics[width=0.5\textwidth]{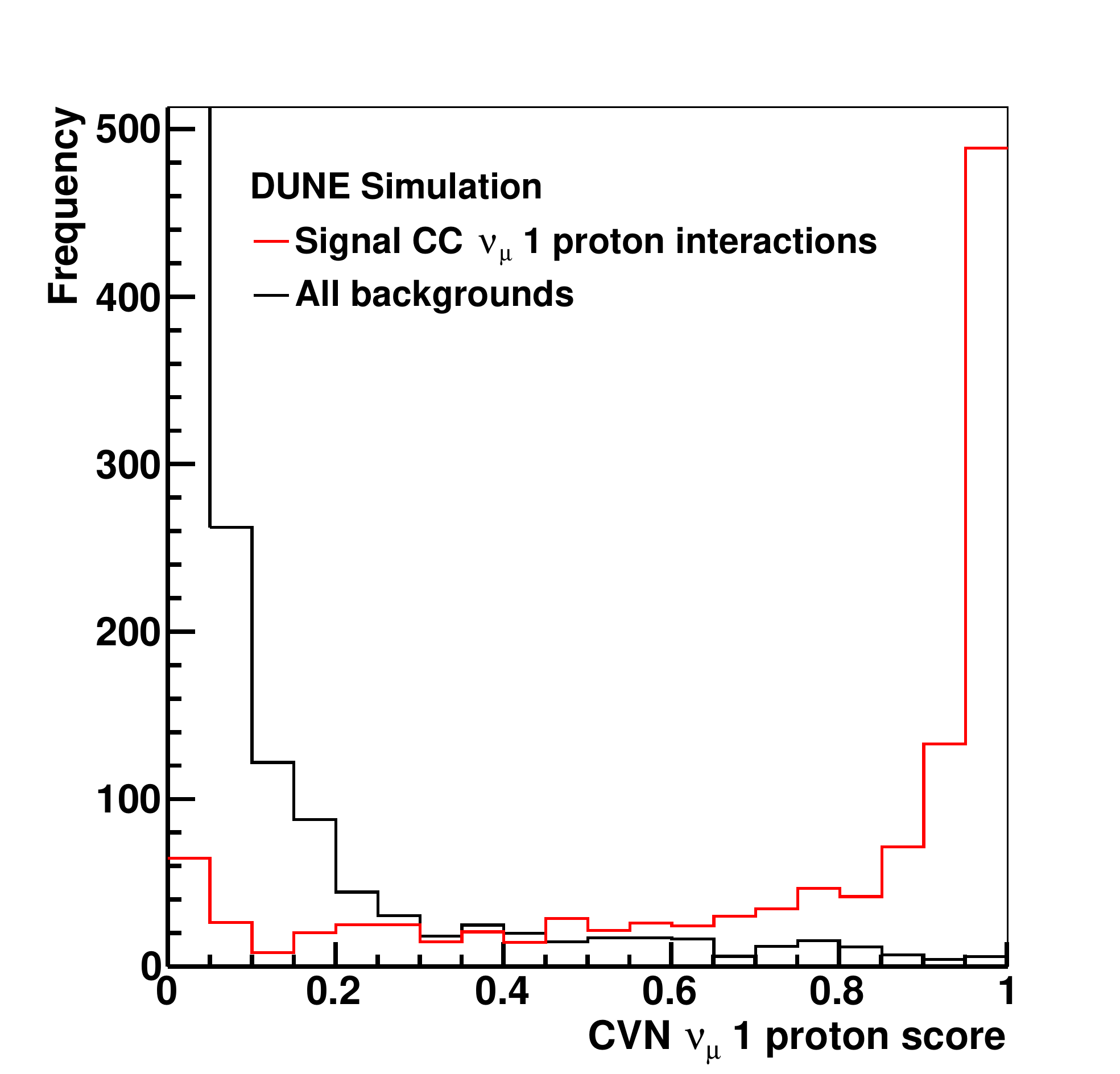}
    \caption{The score for CC $\nu_\mu$ interactions with a single final-state proton from the DUNE CVN. Figure from Ref.~\cite{duneCVN}.}
    \label{fig:dune_cvn_compound}
\end{figure}

Going beyond neutrino flavor classification, the particle counting outputs of the DUNE CVN aim to select interactions with specific final state particles. The output scores for each output are in the range zero to one meaning that a compound score for a given topology can be formed by multiplying the component scores. For example, a score for an event to be a CC $\nu_\mu$ interaction with a single proton in the hadronic final-state system can be written as:
\begin{equation*}
    S(\textrm{CC}~\nu_\mu~1\,\textrm{proton}) = S(\textrm{CC}~\nu_\mu)S(1\,\textrm{proton})S(0\,\pi^\pm)S(0\,\pi^0)
\end{equation*}
Figure~\ref{fig:dune_cvn_compound} shows the distribution of $S(\textrm{CC}~\nu_\mu~1\,\textrm{proton})$ for signal and all background interactions, providing a proof-of-principle for the selection of specific interaction topologies in the DUNE detectors. The ability to sub-divide the neutrino event selections can improve the analysis sensitivity as some events have better energy resolution and lower systematic uncertainties than others.

\paragraph{Event Energy and Position Reconstruction in EXO-200}

EXO-200 is an experiment searching for neutrino-less double beta decay~\cite{Auger_2012}. The detector is a liquid xenon TPC, similar to LArTPC technology, consisting of two drift regions, and each drift volume has two wire readout planes (one induction and one collection) each consisting of 38 wires. Each drift volume also has an array of 37 large-area avalanche photodiodes (APDs) that collect scintillation light. Two different CNN-based algorithms~\cite{Delaquis:2018zqi} have been developed to find the energy and position of candidate events, respectively. The first of these CNNs uses information from the charge readout wires, and the second uses the signals from the APDs.

In the charge-based algorithm, an image is constructed from the charge detected on each of the 76 collection wires in 1024 time samples, resulting in a (1024$\times$76) pixel image. The CNN architecture contains six convolutional layers, each followed by a max pooling layer. A series of three fully connected layers culminates in a single output node that predicts the energy of the interaction. A small improvement, typically a few percent, is seen in the energy resolution at a number of different energies compared to the traditional reconstruction methods.

\begin{figure}
    \centering
    \includegraphics[width=0.95\textwidth]{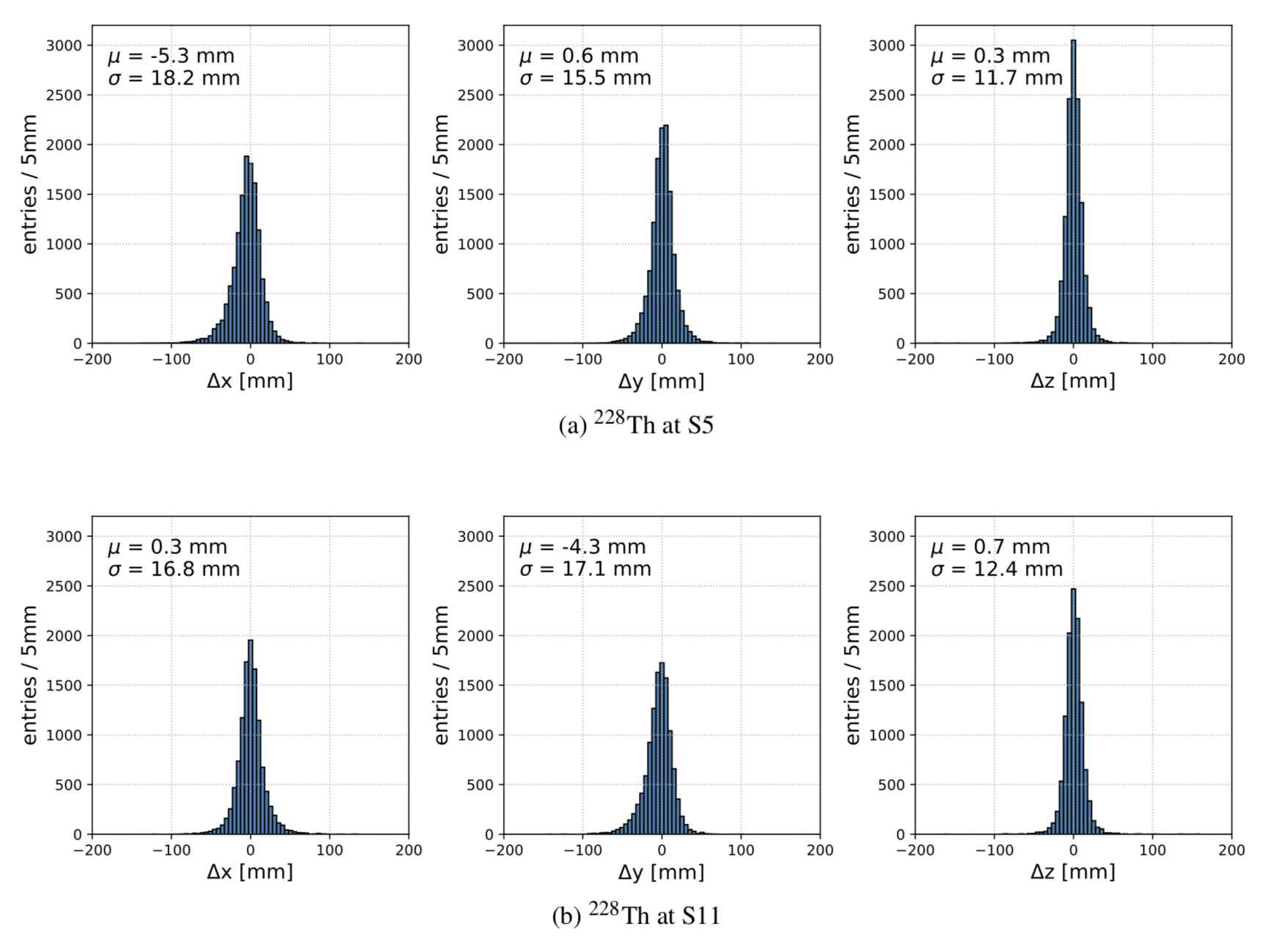}
    \caption{The position residual and resolution for two calibration data samples. The small bias seen in the top left and bottom middle distributions can be attributed to geometric effects. The calibration source for the data shown in a) is located near to the edge of the detector in $x$, and the source used for b) is near the edge of the detector in $y$. Figure reproduced from Ref.~\cite{Delaquis:2018zqi}.}
    \label{fig:exo200_position_resolution}
\end{figure}

The second algorithm aims to reconstruct the position of an event using the distribution of scintillation light detected in the APDs. The (350$\times$74) pixel images are produced from waveforms with 350 time samples for each of the 74 APDs. The CNN architecture chosen consisted of four convolutional layers interspersed with max pooling layers that feed into a three layer MLP, with the final layer returning the $(x,y,z)$ position of the interaction. A novel approach is used to generate the training sample using data labeled with the reconstructed position from the wire readout system, since the charge and light readout systems are independent. This approach will therefore minimize the dependence of CNN performance on different physics models. The data were recorded in a number of calibration runs with different radioactive isotopes. Figure~\ref{fig:exo200_position_resolution} shows that the algorithm works well when applied to the calibration data samples. 

\paragraph{Improving Generalization With Data From the AT-TPC}

The Active-Target Time Projection Chamber (AT-TPC)~\cite{Bradt:2017qgo} is a detector at the National Superconducting Cyclotron Laboratory at Michigan State University.  It is similar to other TPCs discussed here, such as MicroBooNE, DUNE, and EXO-200, but with a few notable differences. Instead of being filled with a liquid noble gas, the AT-TPC is filled with a gas which serves as both the target for the experiment and the drift medium.  The apparatus is placed in a magnetic field so that tracks travel in curved trajectories based on their momentum, and the readout system consists of pads, rather than wires, so that the readout is inherently three dimensional. The AT-TPC is designed to hold a variety of gases to allow for the study of low energy nuclear reactions with low rates.  

Since many experiments run at the AT-TPC, each for a short time and generating large volumes of data, it is particularly important that experiments are able to quickly develop methods for separating signal and background.  Using the $^{46}Ar(p,p)$ experiment, which directed a beam of $^{46}Ar$ ions into the AT-TPC filled with isobutane, a study was performed to determine if CNNs could improve the selection of resonant proton scattering events~\cite{Kuchera:2018djs}.

\begin{figure}
    \centering
    \includegraphics[width=0.8\textwidth]{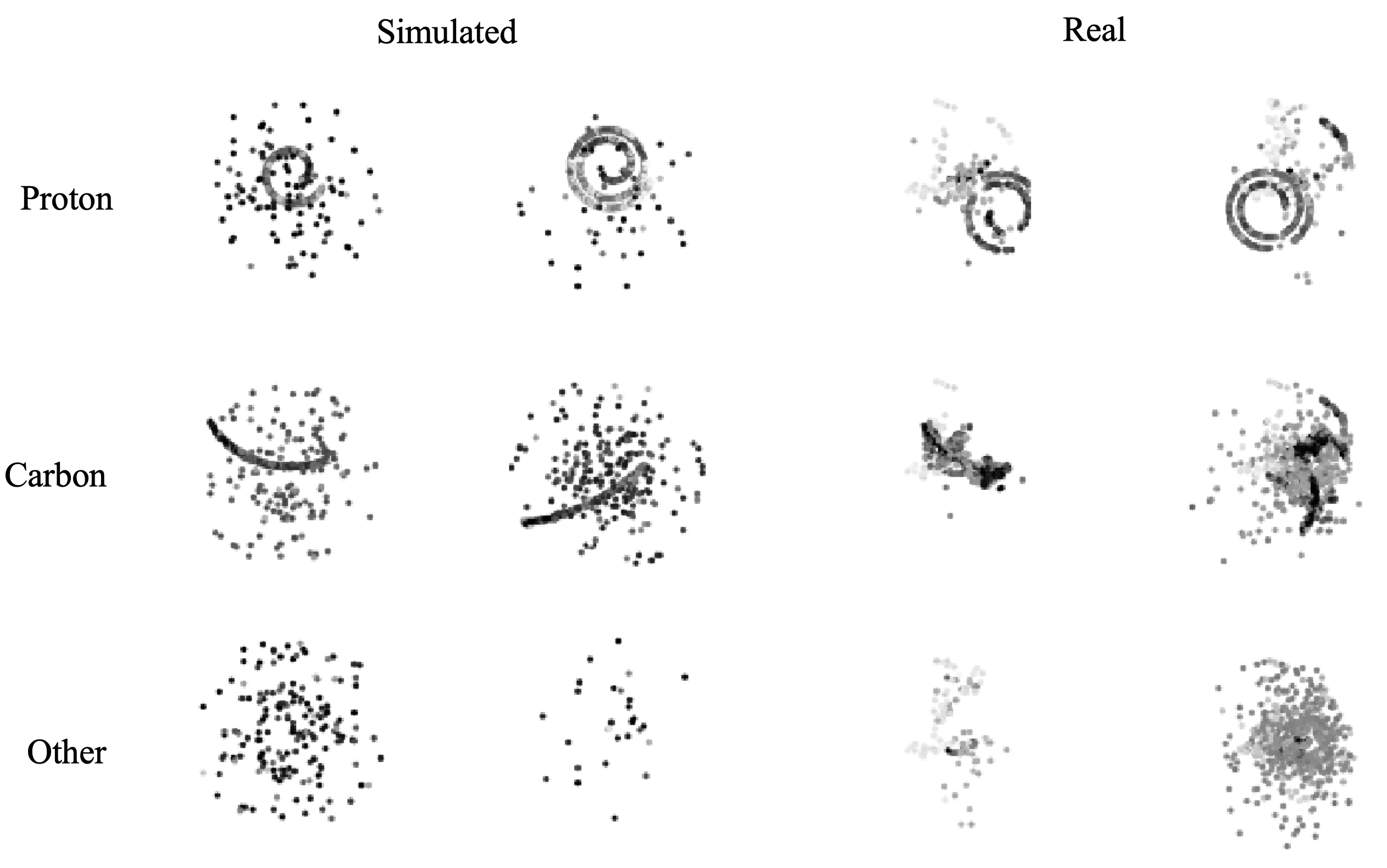}
    \caption{Examples of simulated and real training images for the $^{46}Ar(p,p)$ experiment, where the real training images were extracted through manual classification. The images represent projections of the three-dimensional event data onto the xy-plane. Figure reproduced from Ref~\cite{Kuchera:2018djs}.}
    \label{fig:attpc_data}
\end{figure}

In neutrino and collider experiments previously mentioned, the typically technique is to train a CNN based on a leading architecture using a training sample composed of high-quality simulated data.  Since the available simulations capture many of the important features expected in data, it is typically assumed that this selector will generalize well to experimental data.  This is not a good assumption for the experiments which are conducted at the AT-TPC. For instance, in the $^{46}Ar(p,p)$ experiment, the signal (proton) and one common background (carbon atoms) can be accurately simulated; however, all other backgrounds can not be.  Therefore, CNNs trained with either simulated data, or a small amount of hand-labeled experimental data were studied.  Figure~\ref{fig:attpc_data} shows examples of simulated and real data for proton, carbon, and other categories.

In total, 28,000 simulated training images were produced for each category.  Since manually classifying data is time consuming, only 663 proton, 340 carbon, and 1686 other real training images were produced.  These training sets are too small to be successful using deep CNNs, so they explored the use of transfer learning.  This technique uses the fact that the feature extractor portion of the network is designed to extract low-level features that are properties of images themselves rather than the particular dataset they were trained on while the classifier portion of the network is more tightly tied to the detail of the problem being solved~\cite{10.5555/2969033.2969197}. 

Therefore, a network trained on one dataset may be usable for another dataset after applying a fine tuning procedure.  In this procedure, the classifier portion of the network is removed and replaced with a new one with the correct number of inputs, and the network is retrained using a low learning rate.  In the simplest version of this training, only the weights in the classifier portion are allowed to change.  If the new problem is sufficiently different, it may be necessary to also allow the weights in the feature extractor to change as well.   

To test transfer learning with $^{46}Ar(p,p)$ data, a VGG network~\cite{Simonyan15} previously trained using ImageNet data~\cite{imagenet_cvpr09} was fine tuned using either simulated or real data.  The success of the training was judged using the F1 metric which can be written as

\begin{equation}
    F1 = 2\frac{ \mathrm{precision} \cdot \mathrm{recall}}{ \mathrm{precision} + \mathrm{recall}}
\end{equation}

\noindent where precision is the fraction of true positives out of all positive selected events and recall is the fraction of true positives out of all true events.  In particle physics contexts, precision is often referred to as purity and recall is referred to as efficiency.  When a network that was fine tuned using simulated data was tested on simulated data, the F1 score was $1.0$, signifying perfect classification.  However, when the same network was tested with real data, the F1 score dropped to $0.67$.  This large drop in classification ability is directly related to the low fidelity of the simulated sample.  When a network fine tuned on real data was tested on real data, the F1 score recovered to $0.93$.

This result has a number of interesting implications.  First, the VGG network that was used had been trained on ImageNet data which consists of natural images found on the internet.  Natural images are very different from physics data in that they tend to be information dense while physics data is usually very sparse.  Therefore, it is surprising that transfer learning works at all.  Second, using an exceptionally small sample of manually classified real data (only 663 proton signal examples), it was possible to obtain a selector of similar quality when tested with real data as one trained on simulated data and tested on simulated data.  While neutrino and collider experiments have higher quality simulations than those available for the $^{46}Ar(p,p)$ experiment, there are still concerns about CNNs trained on simulated data increasing the systematic uncertainties of an experiment due to being the network learning the details of the model used in the simulation.  These transfer learning results show that it may be possible to insulate a CNN from such model biases using small quantities of manually classified data.

\section{Convolutional Neural Networks in Heterogeneous Collider Detectors}\label{sec:heterogeneousCases}

Unlike neutrino experiments and other experiments using TPC technology, detectors placed around the collision points of accelerators typically have a cylindrical geometry which the axis of the cylinder aligned with the colliding beams.  Detectors like CMS~\cite{Chatrchyan:2008aa} and ATLAS~\cite{Aad:2008zzm}, located at the Large Hadron Collider are composed of many heterogeneous detector systems organized in concentric layers around the beam axis.  The innermost detectors are usually tracking chambers, consisting of either drift chambers or silicon detectors, designed to measure the trajectory and momentum of charged particles. Placed at larger radii are the electromagnetic calorimeter (ECAL) and hadronic calorimeter (HCAL) systems designed to measure the energy deposited by a variety of particle types.  The outermost layer is designed to identify muons which tend to penetrate much farther than other charged particles. Furthermore,  while these detectors are azimutally symmetric, they have a projective geometry such that detector components are smallest transverse to the beam (at low pseudorapidity) and are largest in very forward regions (at high pseudorapidity).  

Due to these characteristic features, it is more challenging to interpret nearly raw collider data as images.  For any given concentric layer, a popular choice is to unroll the layer at a chosen value of the azimuthal angle so that individual pixels represent bins of pseudorapidity $\eta$ and azimuthal angle $\phi$.  

\paragraph{Quark and gluon jet discrimination}

\begin{figure}
    \centering
    \includegraphics[width=0.9\textwidth]{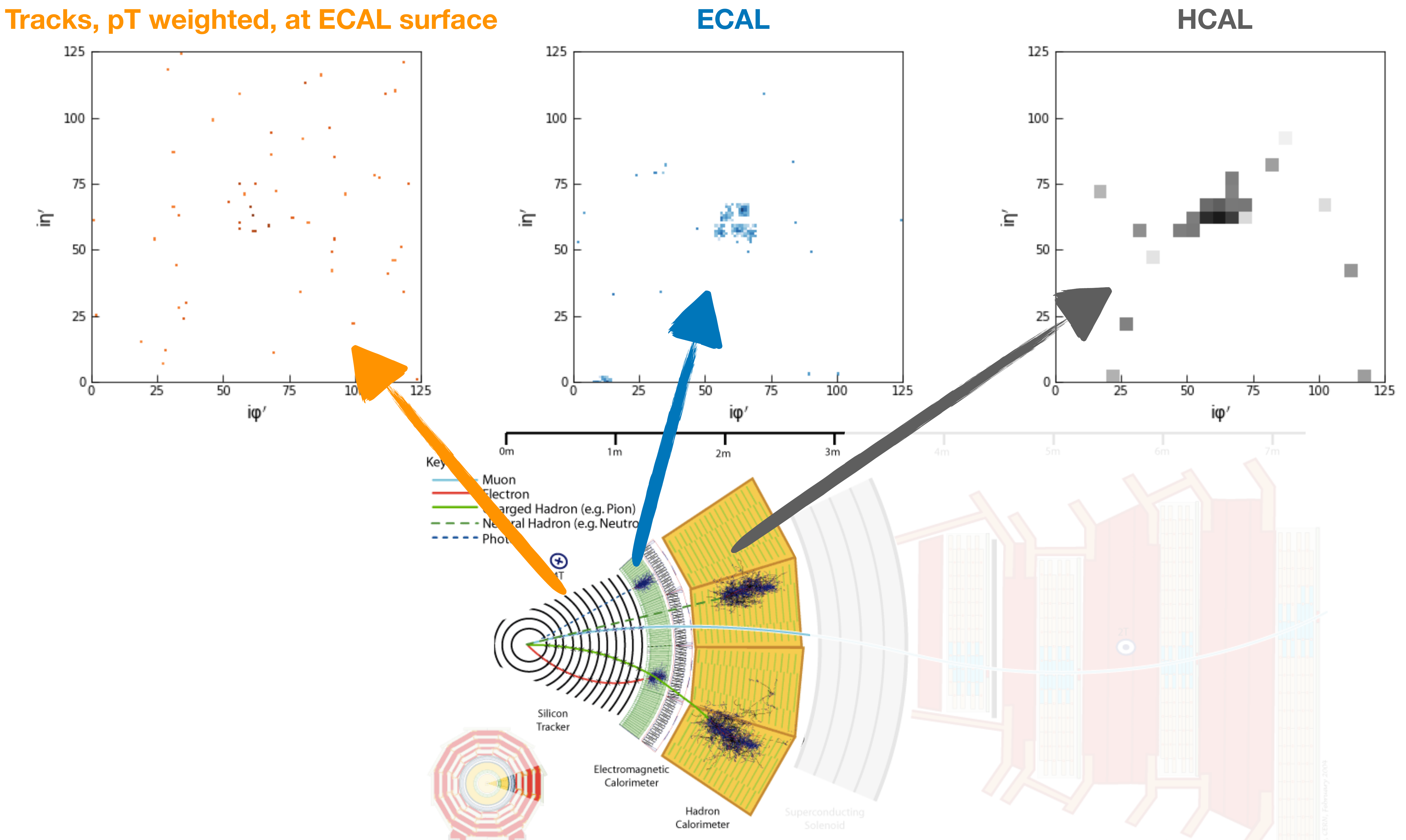}
    \caption{An illustration of the CMS geometry and how to summarize information from the tracking system and the electromagnetic and hadronic calorimeters. Each system has a barrel-shaped geometry which must be recast to form 2D images. Figure reproduced from Ref.~\cite{Alison:2019kud}.}
    \label{fig:CMS_QGDiscrimination}
\end{figure}

A study of quark and gluon jet discrimination using CMS Open Data was presented in Ref.~\cite{Alison:2019kud}. Figure~\ref{fig:CMS_QGDiscrimination} shows how the three images for each event ($p_T$ weighted positions on the front face of the ECAL, and the energy deposits in the ECAL and HCAL, respectively) are produced from the CMS detector geometry. Each of the images are produced using the ($\eta$,$\phi$) coordinate system, with the same binning scheme used for the two ECAL images, and the HCAL has five times coarser images. In order to distinguish between jets initiated by quarks or gluons, an algorithm based on ResNet-15~\cite{He-et-al-2015-deep,He-et-al-2016-identity} was developed that operates on the three images. The results were compared to more traditional techniques using summary information of the reconstructed jets and the CNN approach outperformed them all, achieving an ROC AUC value of $0.8077 \pm 0.0003$ compared to $0.8017 \pm 0.0003$ for the best of the other algorithms.

\paragraph{Electromagnetic shower particle identification} 
Derived variables based on charge and position that describe the shape of energy deposits in collider experiment calorimeters (also known as showers) have traditionally been used to classify showers initiated by different types of particles. In order to distinguish between electron, photon and charged pion showers, the algorithms described in Ref.~\cite{deOliveira:2018lqd} aim to go beyond these variables and use raw data images from the calorimeters.

A six layer MLP neural network operating on 20 shower shape variables provides the baseline algorithm. These 20 variables summarize the information encoded in the raw detector data. Four other networks are considered, the first of which uses the same architecture operating on the 504 calorimeter pixels instead of the 20 variables. Three other networks that operate on three images, one from each layer of the calorimeter, have CNN-based architectures: the locally connected network (LCN)~\cite{deOliveira:2017pjk}, a similar network with the LCN layers replaced by standard 2D convolutions, and a network based on DenseNet~\cite{HuangLW16a}.

\begin{figure}
    \centering
    \includegraphics[width=0.8\textwidth]{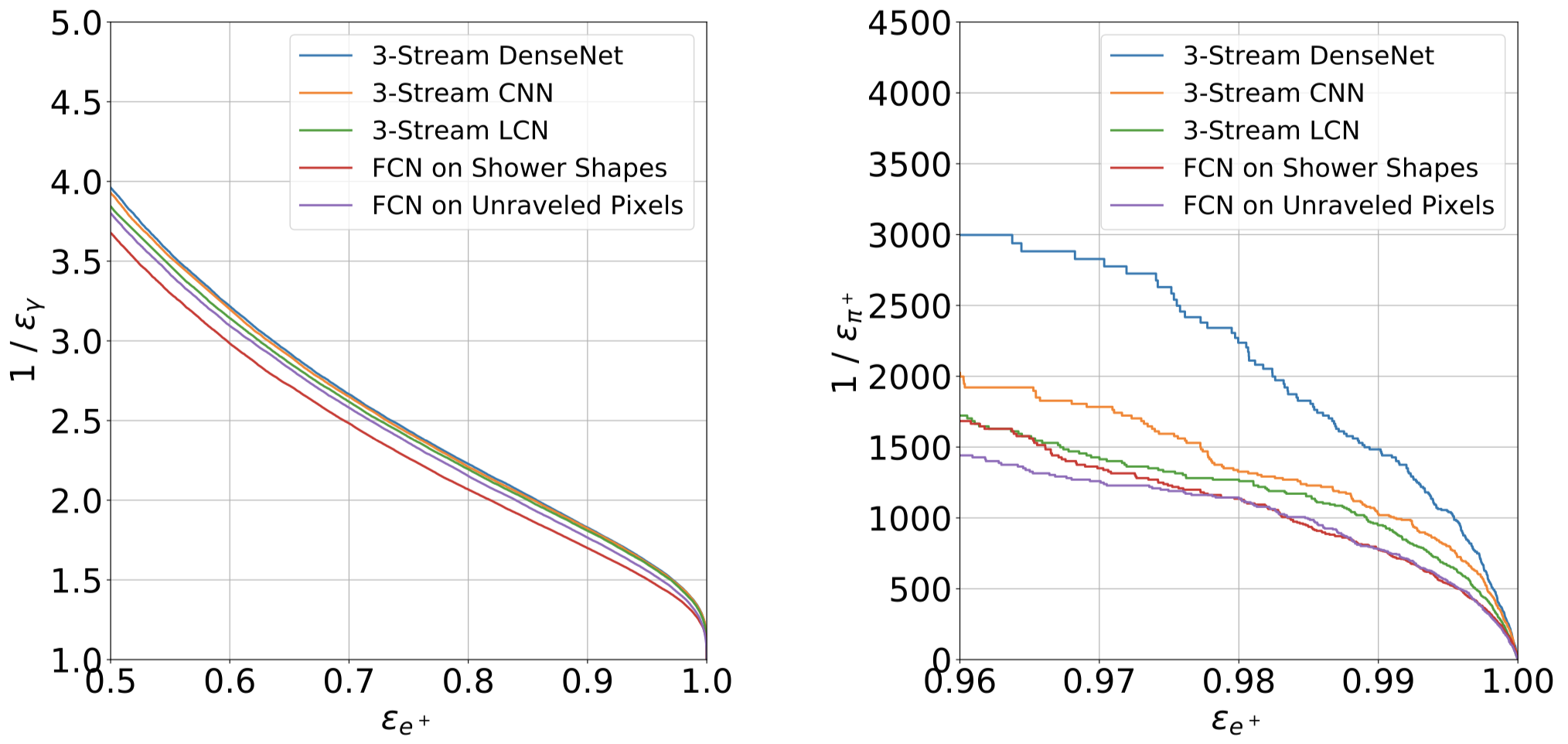}
    \caption{A comparison of five different algorithms in the task of electron-photon (left) and electron-pion (right) shower discrimination. Figure reproduced from Ref.~\cite{deOliveira:2018lqd}.}
    \label{fig:shower_shape_result}
\end{figure}

Figure~\ref{fig:shower_shape_result} shows a comparison of the performance of the five networks for the task of electron-photon separation (left) and electron-pion separation (right). The three CNN-based algorithms outperform the MLP-based ones for both tasks, and the DenseNet-based algorithm demonstrates the best performance overall. It clearly shows that a considerable amount of information is lost in the construction of the shower shape variables and that this extra information is leveraged by the CNNs to significantly improve the performance.

\section{End-to-End Analysis of Time Series using 1D CNNs}\label{sec:1dcnns}

The data produced by a single sensor is generally a continuous waveform: a signal that varies as a function of time. In many cases this 1D representation of data needs to be processed without combining data from multiple sensors to form image, for example, when there is a need for pre-processing of the data from each individual sensor, or when the number of sensors is small and each will be processed individually. 

A natural way to process these 1D waveforms is to use 1D convolutions. The $n$-element filters are applied to the input waveform to extract features in an analogous way to the extraction of image features in the 2D case as previously discussed. For example, a 1D convolution algorithm could be used to find peaks in a waveform to find energy deposits recorded by a given sensor. Other neural networks, such as recurrent neural networks (RNNs)~\cite{rnn_little,rnn_hopfield} and their subclass long short-term memory networks (LSTMs)~\cite{lstm}, can be used on 1D waveforms, but 1D CNNs work very well on fixed length inputs such as those from detector elements with a fixed-length readout window. An example of event classification using a 1D CNN is given below.


\paragraph{Pulse Shape Discrimination for Scintillation Signals}

\begin{figure}
    \centering
    \includegraphics[width=0.5\textwidth]{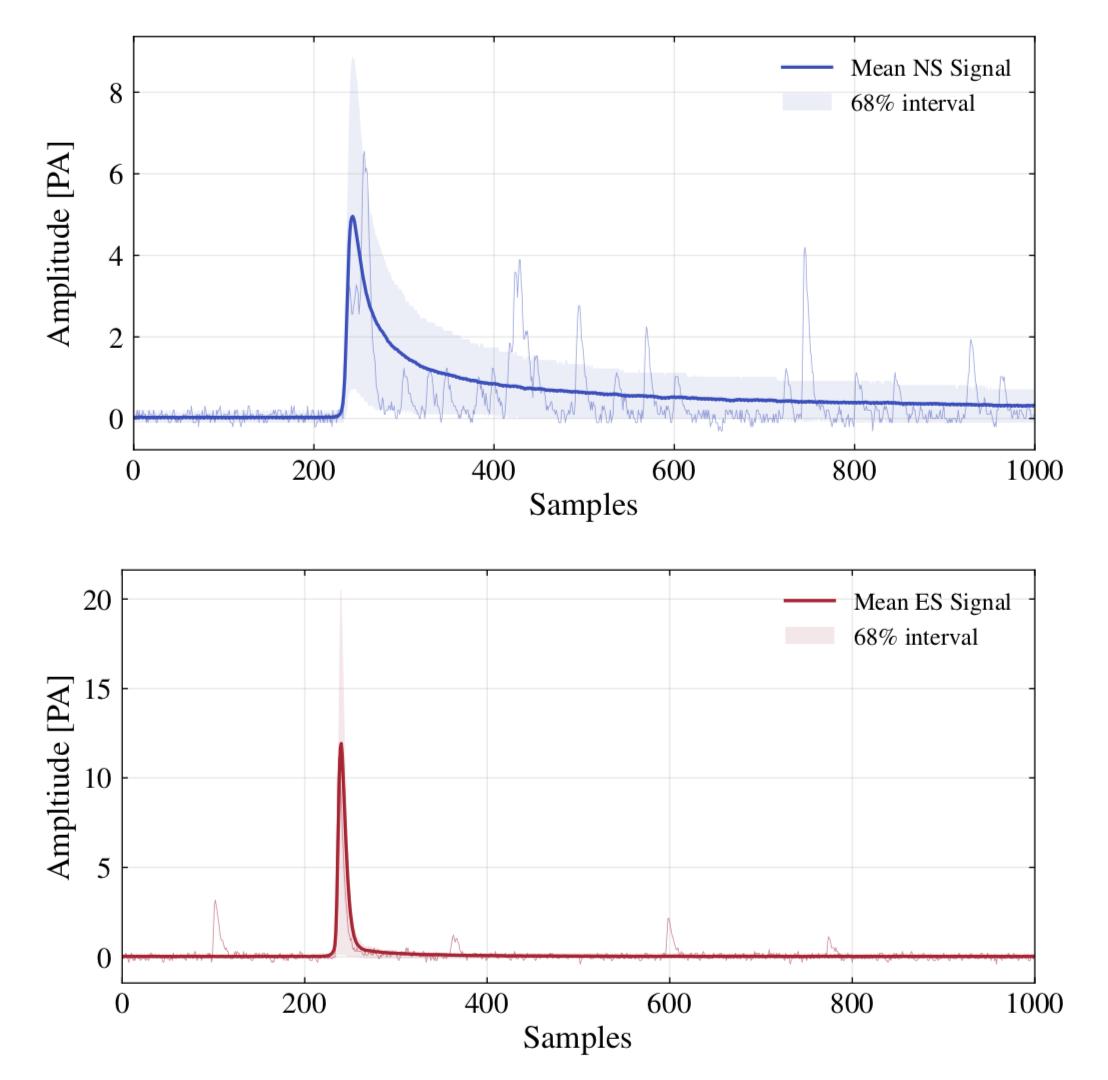}
    \caption{The average waveform for NS (top) and ES (bottom) events where the shaded regions show the 68\% interval of the ensemble used to calculate the average. Also shown in both panels is an example waveform. Figure reproduced from Ref.~\cite{Griffiths:2018zde}.}
    \label{fig:psd_waveform}
\end{figure}

Pulse shape discrimination, the ability to identify different signals in raw waveforms, is a common task in high energy physics. In this example~\cite{Griffiths:2018zde}, the experimental setup consists of a $^6$LiF:ZnS(Ag) phosphor screen coupled to a scintillator cube, technology similar to that used in the SoLiD experiment~\cite{Abreu:2017bpe}. The light produced inside the scintillator cube was read out using a photomultiplier tube (PMT) and two silicon photomultipliers (SiPM). The scintillator cube was sensitive to interactions from gamma-rays and electrons that produce scintillation light signals referred to as electron scintillation (ES). The phosphor screen was sensitive to nuclear interactions producing a different light signal, referred to as nuclear scintillation (NS). The goal of the experiment was to distinguish between the ES and NS events from the raw SiPM waveforms, where Figure~\ref{fig:psd_waveform} shows the average NS (top) and ES (bottom) waveforms. The PMT served two purposes: to trigger the readout of the SiPM waveforms, and to label the waveforms, with approximately 99\% accuracy, as either an ES or NS event to avoid the need for simulations. Each of the SiPM waveforms consisted of 1000 samples.

\begin{figure}
    \centering
    \includegraphics[width=0.7\textwidth]{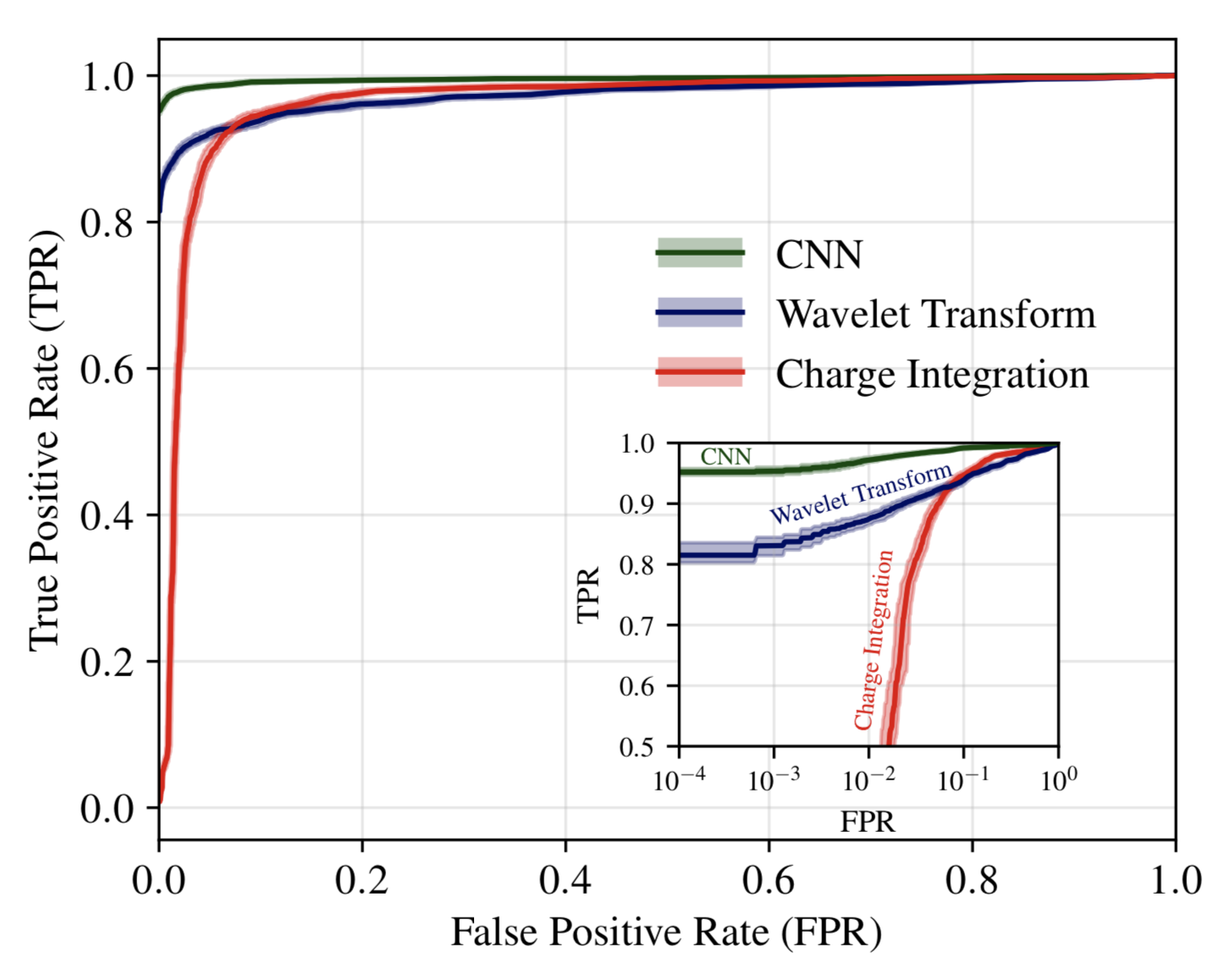}
    \caption{ROCs curves for the different PSD algorithms. Figure reproduced from Ref.~\cite{Griffiths:2018zde}.}
    \label{fig:psd_1DCNN}
\end{figure}

A 1D CNN was developed using just two convolutional layers, each followed by a max pooling layer. The output of the second pooling layer fed into a fully connected layer, and finally a single output node with a softmax activation to provide the probability of the waveform being of the NS type. The CNN algorithm significantly outperforms two more traditional approaches, as demonstrated by the distributions in Figure~\ref{fig:psd_1DCNN}. 

\section{Graph Neural Networks for Large Three-Dimensional Detectors}\label{sec:gnnCases}

Graph neural networks are a more recent development than CNNs, and as such GNNs are currently less commonly used in high-energy physics than CNNs. However, GNNs are beginning to be used in event reconstruction~\cite{farrell2018novel,Qasim_2019,ju2020graph}, as described in detail in Chapter~12. It is likely only a matter of time before many examples of end-to-end analysis using GNNs become apparent, but the only current example is discussed below.

\paragraph{Event classification in the IceCube experiment}
IceCube~\cite{Aartsen:2016nxy} is an experiment located in Antarctica that aims to measure interactions of atmospheric and astrophysical neutrinos. The detector consists of a series of photomultiplier tube detector modules (called DOMs) buried in the ice to measure Cherenkov radiation produced by charged particles traveling in the ice. There are approximately 6000 DOMs arranged in an irregular 3D hexagonal geometry making it well-suited to graph representation. Each of the DOMs is represented as a graph node with six features: the $\left(x,y,z\right)$ position, the sum of the charge in the first detected pulse, the sum of the charges from all pulses, and the time at which the first pulse went above threshold. On an event-by-event basis, only those DOMs that record a signal are added as nodes to the graph. The goal of the GNN is to classify an event (i.e. the graph as a whole) as either a signal neutrino interaction or a background event in an environment where the signal interactions are very rare compared to the backgrounds. The network architecture is based on the MoNet model~\cite{monti2016geometric}.

\begin{figure}
    \centering
    \includegraphics[width=0.8\textwidth]{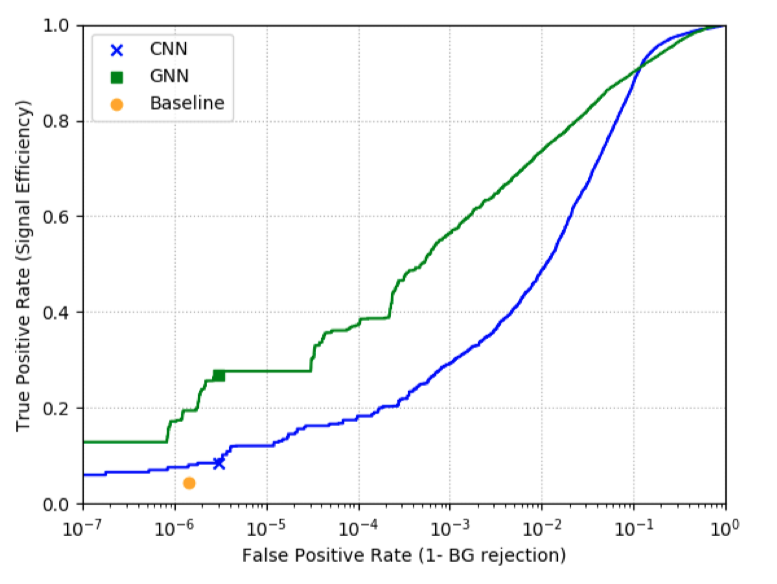}
    \caption{Distribution showing the signal efficiency as a function of the false positive rate. The IceCube GNN algorithm (green) is compared to a 3D CNN (blue) and a baseline point (yellow) from traditional techniques. Figure reproduced from Ref~\cite{icecubeGCN}.}
    \label{fig:icecube_gnn_roc}
\end{figure}

The local neighborhood of each node, or its adjacency to other nodes, is defined using the three spatial position features $\left(x,y,z\right)$. The edges between nodes are assigned weights from a Gaussian distribution that depends on the distance between the two nodes. The width of this Gaussian distribution is a learned network parameter that controls how quickly information is spread between spatially distant nodes. A series of convolutions are applied to the graph followed by a logistic regression to predict the event type. Figure~\ref{fig:icecube_gnn_roc} shows the true positive rate as a function of the false positive rate that the GNN significantly outperforms traditional methods as well as 3D CNN approaches, achieving a signal-to-noise ratio of 2.98 compared to the baseline of 0.987~\cite{icecubeGCN}.

\section{Opening the Black-Box} \label{sec:networkbehaviour}

In traditional selection techniques, reconstructed features are designed to quantify a physical property known to differ between event types.  For instance, the multiple scattering distribution for muons and charged pions differ because muons only scatter due to the Coulomb potential of the material they are propagating through while charged pions also scatter due to the strong nuclear potential.  Therefore, it is reasonable to expect that summary statistics like the mean and RMS of the scattering angles of a track may be useful when classifying tracks as having been created by a muon or charged pion. 

Since end-to-end approaches use more information than these summary statistics, they typically perform better, but it is more difficult to attribute their performance to understandable physical properties. Moreover, using more information potentially exposes the algorithm to learning spurious or incorrect details due to imperfections in the simulated training dataset.  Therefore, it is critical to have tools for interrogating the network to determine on what basis it is making its decisions. For a final physics analysis, this involves a black-box input-output analysis where systematically-varied simulation samples are classified by the network to determine how sensitive the classification is to plausible variations in the dataset. However, tools which provide a qualitative understanding the network's decisions can provide additional assurance of reasonableness. In this section, we will discuss examining feature maps at various depths in the network to identify features frequently associated with certain classifications, low-dimensional visualizations of the features produced by the final layer of the network to determine how a test sample forms clusters, and occlusion tests to determine what parts of an image are more salient for making a decision. In all of the cases given below, the methods produce figures that are inspected by eye to determine if the behavior appears reasonable; there is no unique quantitative figure-of-merit that works for all networks.

\subsection{Feature maps}

The outputs of the convolutional layers can provide insight as to what features are being extracted from the input images. Looking at the output of the first convolutional layer for different types of events can show what sort of features the network is looking for in order to classify the event. However, the layer outputs become increasingly abstract as the depth into the network increases making visual inspection of the deeper layer outputs difficult. 

\begin{figure}
    \centering
    \includegraphics[width=0.7\textwidth]{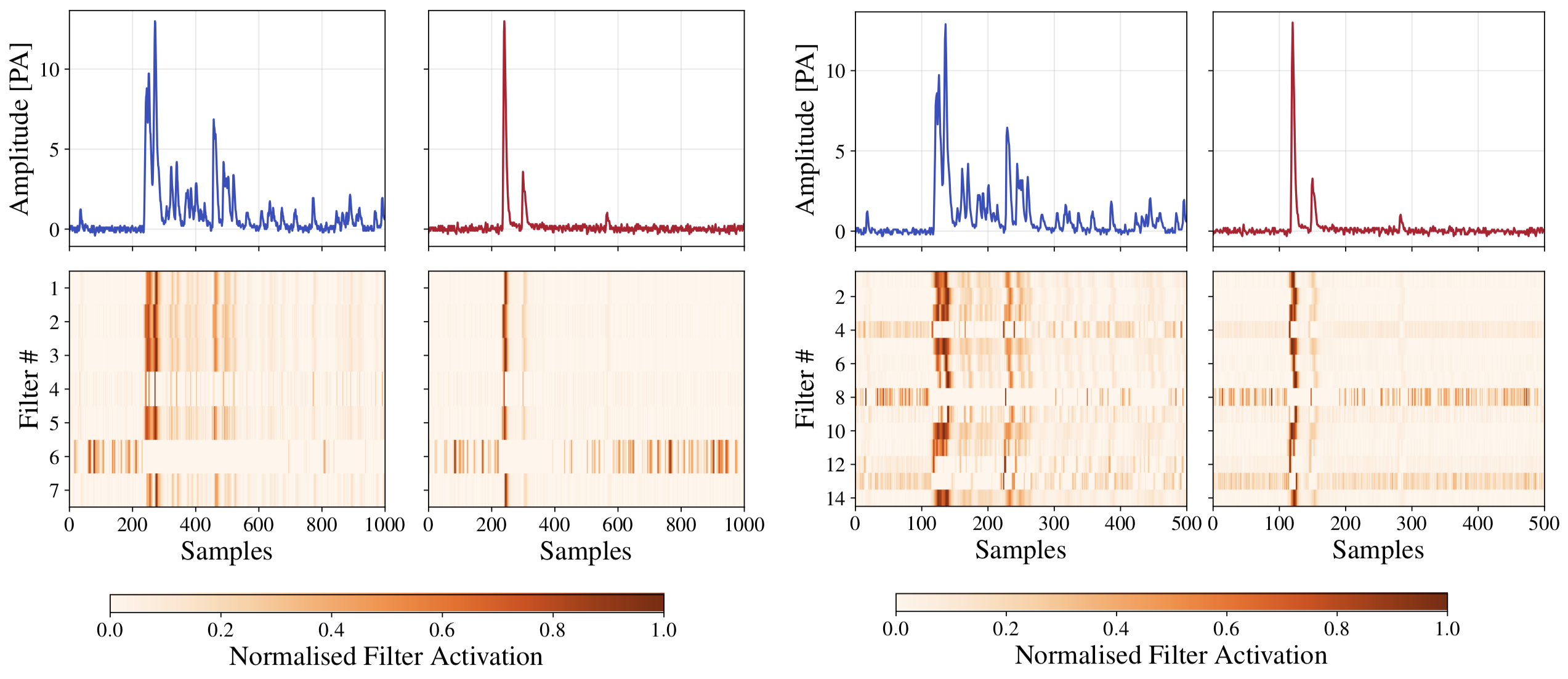}
    \caption{Normalized feature outputs of the first (left) and second (right) convolutional layers for the two types of signal waveforms. The input signals on the right have been downsampled (to mimic the max pooling in the CNN architecture) to show the correlation with the second convolutional layer. Figure reproduced from Ref.~\cite{Griffiths:2018zde}.}
    \label{fig:pds_featuremap}
\end{figure}

Figure~\ref{fig:pds_featuremap} shows example feature maps from the 1D CNN described in Ref.~\cite{Griffiths:2018zde} and Section~\ref{sec:1dcnns}, for the first convolutional (left) and second (right) convolutional layers for the two types of signals. For example, visual inspection of the distributions on the left shows that filter six responds most to low amplitude samples and filters 2 and 3 find the main peaks in the waveforms.

\begin{figure}[!htbp]
    \centering
    \includegraphics[width=0.95\textwidth]{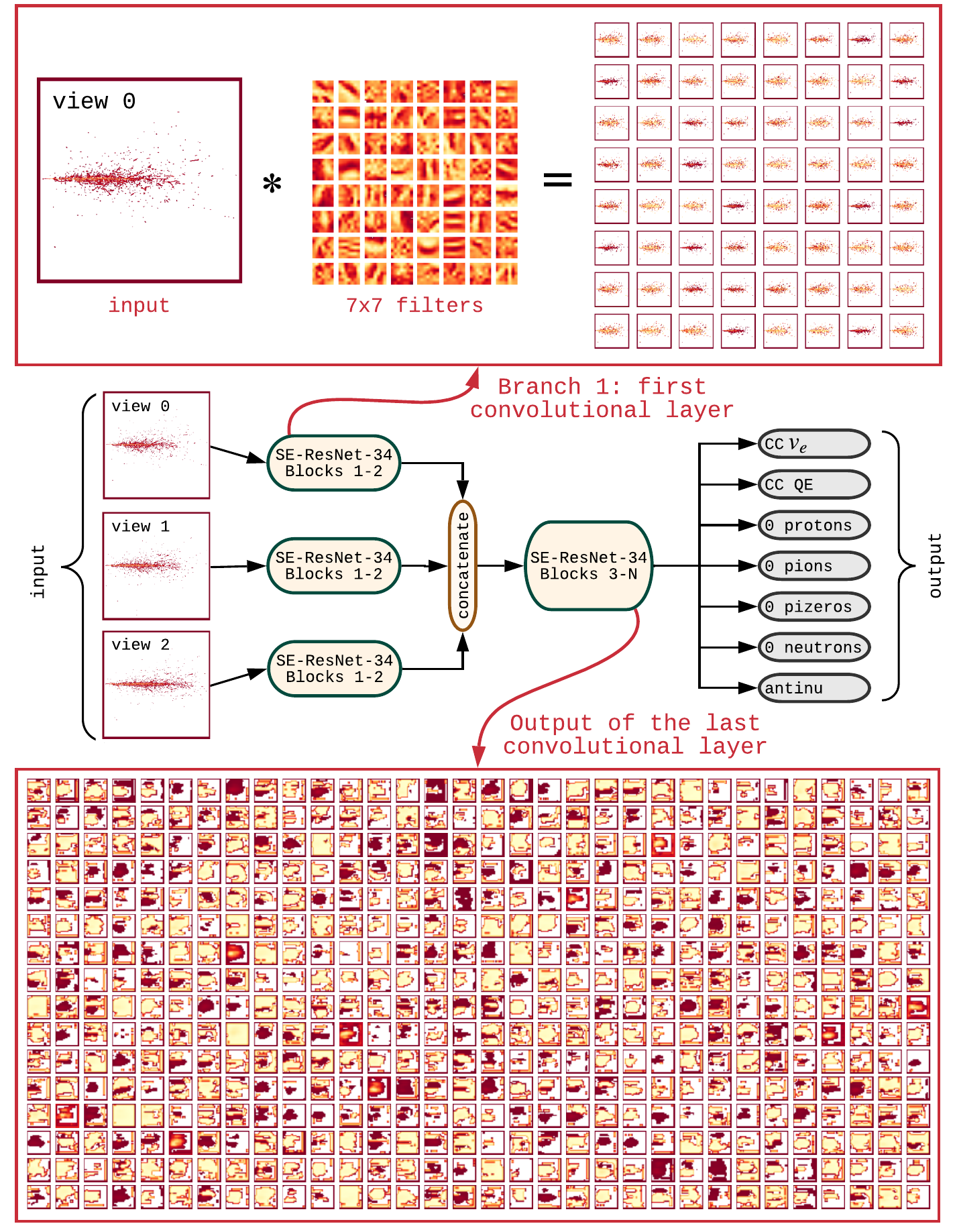}
    \caption{The outputs from the first (top right) and last (bottom) convolutional layers in the DUNE CVN for an input CC $\bar{\nu}_e$ interaction. The response strength is shown ranging from yellow for low response values to red for a strong response, and white is used for no response due to empty input pixels. Figure reproduced from Ref~\cite{duneCVN}.}
    \label{fig:dune_feature_maps}
\end{figure}

As an example from a 2D CNN, Figure~\ref{fig:dune_feature_maps} shows the response from the first and final layers of the DUNE CVN~\cite{duneCVN} for an input CC $\bar{\nu}_e$ interaction. The first convolutional layer consists of 64 learned $7\times7$ pixel filters, visualized in the top middle panel. The results of applying these 64 filters to the input image are shown on in the top right panel, demonstrating that some filters result in a weak (yellow) response and some give a strong (red) response for the chosen input image. Some of the filters can be seen to respond strongly to the central part of the shower, others to the sparser halo pixels, and some have a weak response for the whole image. The 512 feature maps from the final convolutional layer, shown in the bottom panel, have a large variety in response but are very abstracted since the many pooling layers have downsampled the original $500\times500$ pixel input down to $16\times16$ pixels in the feature maps. 

\subsection{Low-Dimensional Visualizations}

For both CNNs and GNNs discussed in this chapter, the network consists of a feature-extractor and classification sub-network trained simultaneously. In the CNN case, the feature extractor consists of a stack of convolutional layers, with the possibility of pooling or other types of layers.  In the GNN case, the feature extractor typically involved convolutions generalized to the graph structure and message passing steps.  In either case,  the feature extractor produces a high-dimensional vector encoding of the information contained within the input data. The classification sub-network usually consists of a single fully connected layer with as many output nodes as classes and the outputs recast as class probabilities using the softmax function. 

In the simplest case, predicted outputs are merely rescaled linear combinations of the output of the feature extractor making decision boundaries in the high-dimensional feature space encoding hyper-planes.  Therefore, in a successfully trained network, examples from the same class should be separated by a small Euclidean distance, while examples from different classes should be separated by a large Euclidean distance. 

Examining the shape of clusters within the high-dimensional feature space could provide insights into why different examples are classified correctly or incorrectly.  For instance, two classes may exist as to well-separated clusters except at one surface where they touch.  At the point of contact, the proper classification of those examples would be ambiguous.  This makes it possible to isolate only those examples with a true ambiguity. Unfortunately, the output of the feature extractor typically has a dimensionality of $\mathcal{O}(1000)$, well beyond the bounds of normal visualization techniques.  

The t-distributed stochastic neighbor embedding (t-SNE) technique~\cite{tsne}, provides a method to visualize high dimensional data by embedding it in a lower dimensional space.  To do this, the similarity representing the probability that two points are neighbors is constructed for each pair of points in either the high or low dimensional space as a function of the Euclidean distance between the points. In the high dimensional space, the similarity is based on a Gaussian probabilty density while it is based on Student's t-distribution in the low dimensional space.  The Student's t-distribution prevents points from crowding in the low dimensional space.  Finally, the points in the low dimensional space are rearranged to minimize the Kullbach-Leibler divergence between the similarity distributions in both representations assuring that points have the same relationship to their neighbors in low dimensions as in high. 

\begin{figure}
    \centering
    \includegraphics[width=0.9\textwidth]{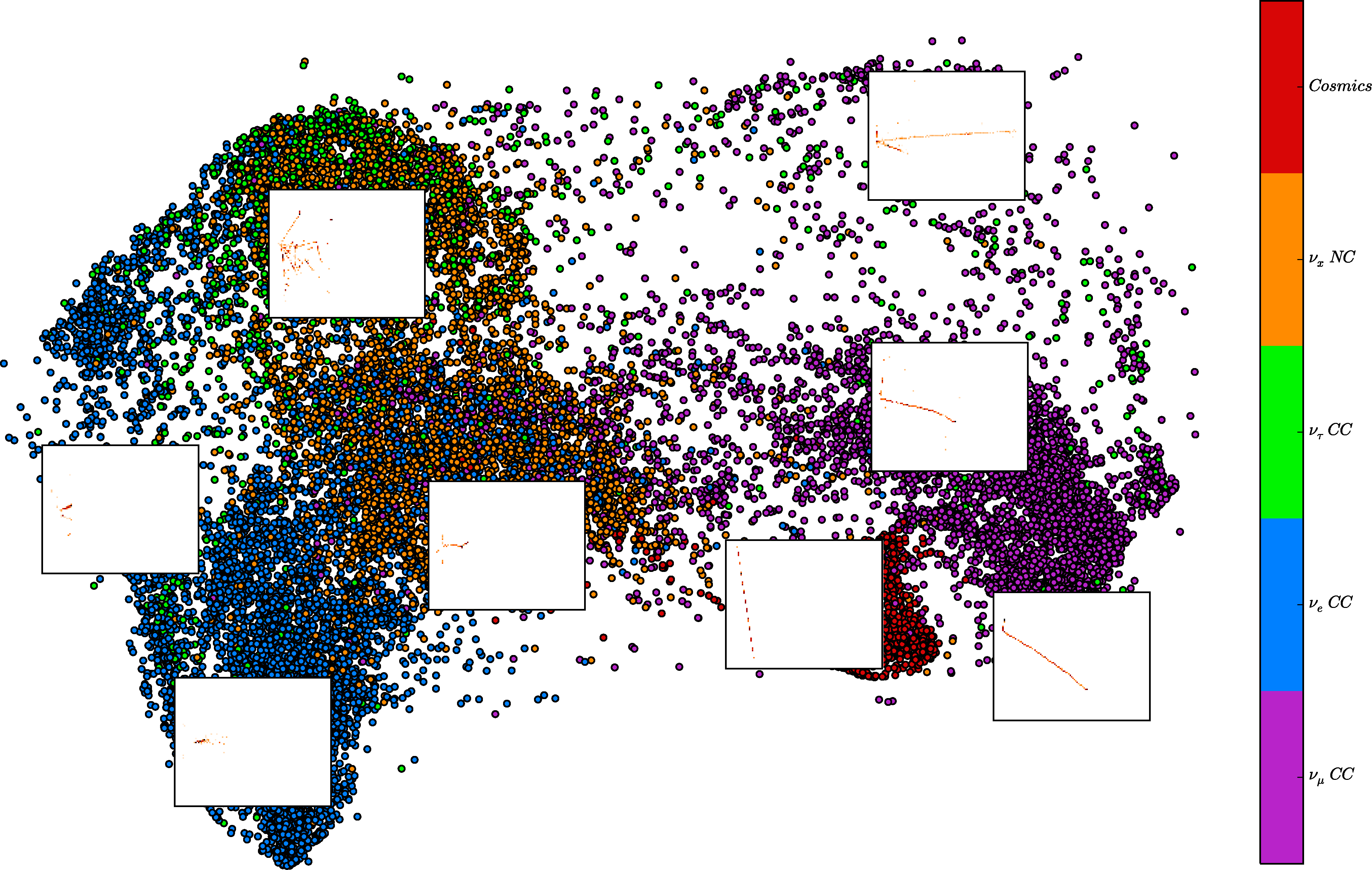}
    \caption{A visualization of the t-SNE algorithm applied to the 1024 dimensional outputs of the NOvA CNN~\cite{novacvn}. Figure courtesy of the NOvA collaboration.}
    \label{fig:nova_tsne}
\end{figure}

Figure~\ref{fig:nova_tsne} shows t-SNE algorithm~\cite{tsne} applied to 1024 dimensional outputs of the NOvA CNN~\cite{novacvn}. The structure of the clusters gives insight into what features the network primarily used to separate neutrino interactions. Looking at visualizations of a  representative sample of interactions, it is clear what characteristics define the axes of the low-dimensional representation.  On the horizontal axis, events become dominated by a long track on the right side and dominated by an electromagnetic shower on the left side.  On the vertical axis, the multiplicity of objects in the interaction increases while moving from the bottom to the top of the figure.

With this scheme in hand, we can understand how the clusters of classes are arranged. We see that $\nu_{\mu}$ are well separated from $\nu_{e}$ with neutral current interactions in the middle.  This makes sense since $\nu_{\mu}$ interactions are dominated by tracks while $\nu_{e}$ interactions are dominated by an electromagnetic shower.  Neutral current interactions lack a charged lepton, but they can contain either a charged pion (track-like) or a neutral pion (shower-like) which naturally causes ambiguities with either $\nu_{\mu}$ or $\nu_{e}$.  Cosmic rays largely occupy the lower right side of the figure since they are dominated by single muons.  Finally, $\nu_{\tau}$ interactions are clustered in the top half of the figure since they only occur at higher energies, and they are poorly separated from the three other neutrino channels since the $\tau$-lepton can decay either hadronically or leptonically. In broad terms, this figure suggests that the network is relying on similar topological features to what a hand scanner might use.  Furthermore, it suggests potential improvements to the network.  For instance, since cosmic rays consisting of single muons are easy to separate, and the selector may be improved by constructing a biased sample of cosmic rays emphasizing more complicated topologies which are rarer, but more likely to be mis-categorized as a neutrino.  Similarly, maintaining a single category for $\nu_{\tau}$, regardless of the decay mode of the $\tau$-lepton needlessly confuses the network. 

\subsection{Occlusion tests}

While t-SNE provides general insight into what type of examples are seen as similar by the network, it can be useful to determine precisely which portions of an example are salient to the decision made by the network.  A very simple method for determining salience is the occlusion test.  In an occlusion test, a small portion of an input example is withheld from the network.  In the CNN case, this would mean changing a small patch of pixels to zeros.  The change in the network output is placed into a separate map  at the pixel corresponding to the center of the occluded region in the input image.  Repeating this across the image produces a salience map showing which regions were most important in making a particular decision. 

\begin{figure}
    \centering
    \begin{tabular}{cc}
        \includegraphics[width=0.47\textwidth]{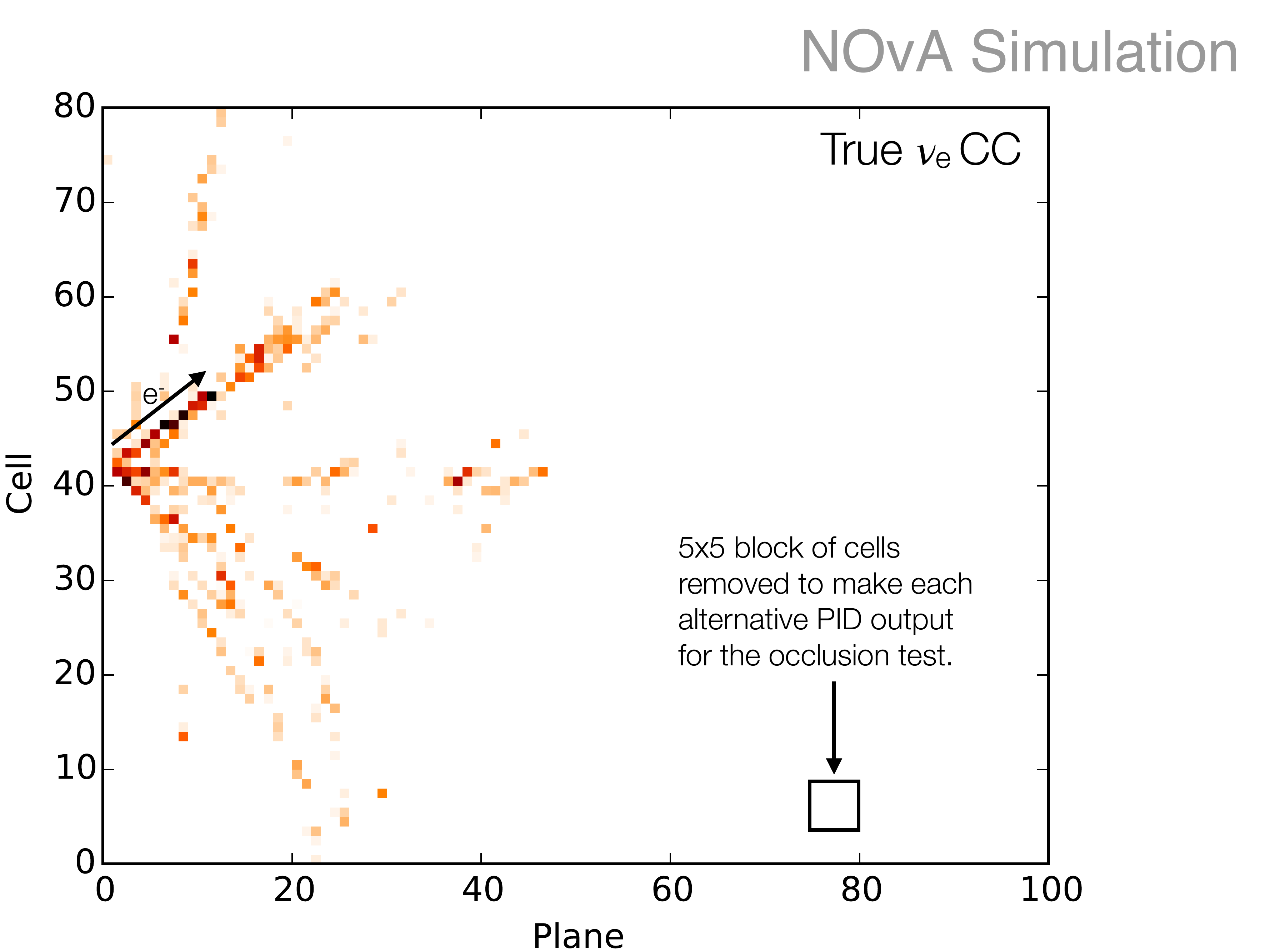} &
        \includegraphics[width=0.47\textwidth]{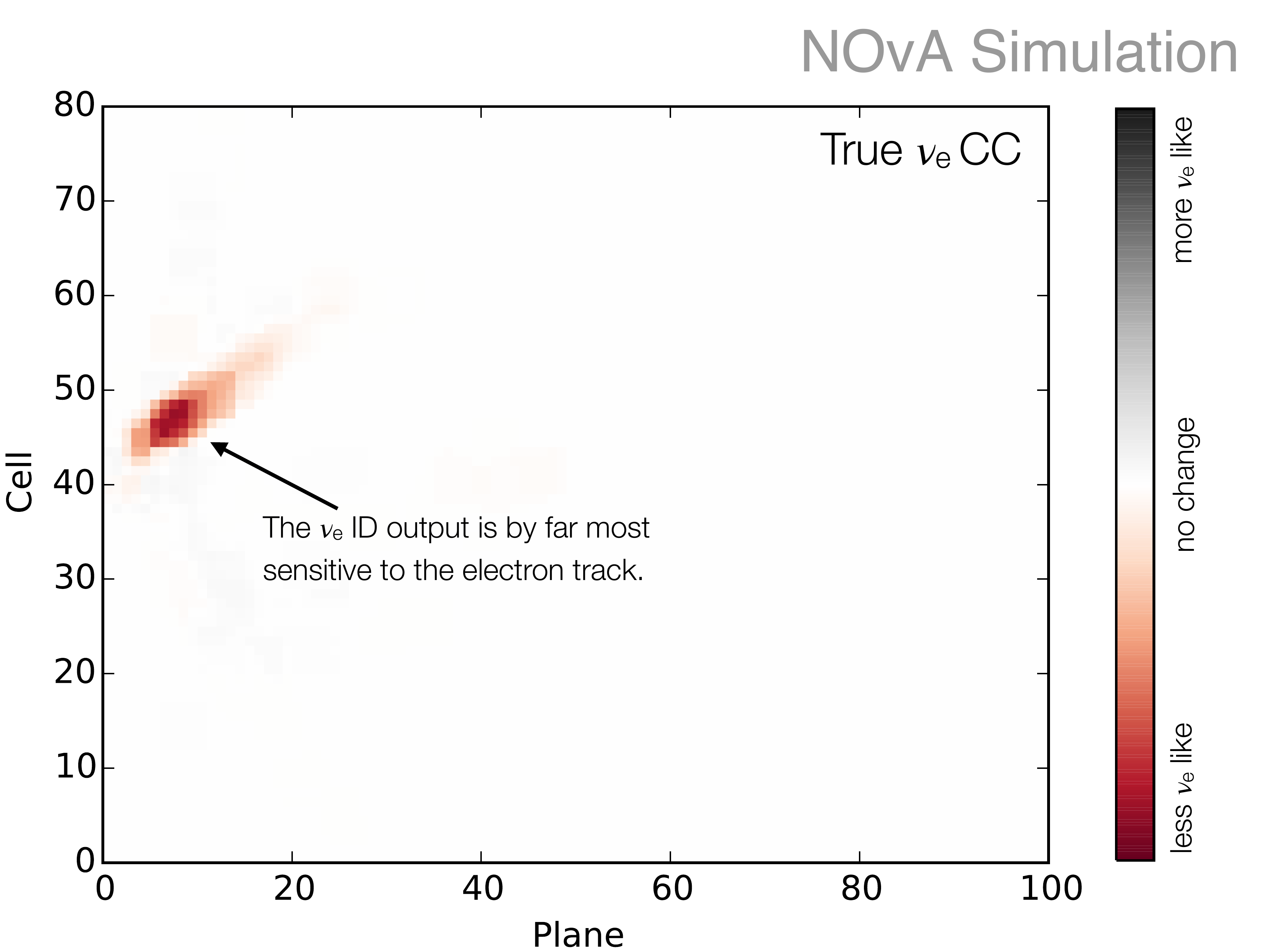}
    \end{tabular}
    \caption{An occlusion test demonstrating the most salient parts of an input image to the decision made by the NOvA CNN~\cite{novacvn}.  (left) A single view of a true $\nu_{e}$-CC interaction. This interaction consists of a single electromagnetic shower from a primary electron along with several track-like objects from the hadrons produced by the nucleus. (right) The change in the $\nu_{e}$-CC score as a function of the location of a $5\times5$ occluded region. Figures courtesy of the NOvA collaboration.
    \label{fig:nova_occlusion}}
\end{figure}

Figure~\ref{fig:nova_occlusion} shows the salience map for a deep inelastic scattering $\nu_{e}$ interaction using the NOvA CNN~\cite{novacvn}.  In this case, the occluded region consisted of a movable $5 \times 5$ square of pixels.  Since the image shown corresponds to a deep inelastic scatter,  the interaction consists of an electron produced by the charged current interaction and an array of charged and neutral pions produced by the struck nucleus.  Despite the high multiplicity of this interaction, the only region which reduces the $\nu_{e}$ score of this interactions is near the start of the electron shower.  This is consistent with how a hand scanner would classify this event since electrons produce a single track before they initiate a shower while photons are invisible until they initiate their first pair production.  Furthermore, the lack of dependence on the details of the particles produced by the struck nucleus is reassuring since that portion of the simulation is typically less realistic.

\section{Conclusions}\label{sec:conclusion}

End-to-end analyses using deep learning are becoming widespread in high energy physics, taking raw detector data as input and providing physics-level outputs such as event classification. The majority of algorithms to date are based on 2D convolutional neural networks applied to images of the experimental detector data, but other approaches such as graph neural networks are gaining popularity. The examples presented here demonstrate that end-to-end deep learning analysis are very powerful because they have access to all of the detector information and they can significantly outperform more traditional analysis techniques.

As with any type of analysis, it is important to ensure robustness and to understand how the event classification is being performed. The techniques outlined in Section~\ref{sec:networkbehaviour} help to elucidate how the deep neural networks extract features from the input data and use the features to perform separation of the different categories within some high-dimensional space.

\bibliographystyle{tepml}
\bibliography{mainbib}


\end{document}